\documentclass{WileyChemistry-template}

\usepackage{graphicx}
\usepackage{dcolumn}
\usepackage{bm}
\usepackage{makecell}
\usepackage{multirow}
\usepackage{siunitx}
\usepackage{comment}
\usepackage{numprint}
\usepackage{longtable}
\usepackage{tabularray}
\usepackage{numprint}
\usepackage{booktabs}
\DeclareSIUnit\bar{bar}
\usepackage{soul}
\usepackage{placeins}

\usepackage{ulem}
\DeclareSIUnit\electronvolt{eV}
\DeclareSIUnit\Atomicunit{a.u.}

\newcommand{\etal}{\textit{et al.}}

\title{\raggedright Excited state assignment and state-resolved photoelectron circular dichroism in chalcogen-substituted fenchones}

\author{
	\begin{minipage}{\textwidth}
		Sudheendran Vasudevan,\textsuperscript{[a]} 
		Steffen M. Giesen,\textsuperscript{[c]} 
		Simon T. Ranecky,\textsuperscript{[a]}
		Lutz Marder,\textsuperscript{[a]}
		Igor Vidanovi\'{c},\textsuperscript{[d]}
            Manjinder Kour,\textsuperscript{[c]} 
            Catmarna K\"ustner-Wetekam,\textsuperscript{[a]}
		Nicolas Ladda,\textsuperscript{[a]}
		Sagnik Das,\textsuperscript{[a]}
		Tonio Rosen,\textsuperscript{[a]} 
            Vidana Popkova,\textsuperscript{[a]}
		Hangyeol Lee,\textsuperscript{[a]} 
		Denis Kargin,\textsuperscript{[d]}  
		Tim Sch\"afer,\textsuperscript{[e]}
            Andreas Hans,\textsuperscript{[a]}
		Thomas Baumert,\textsuperscript{[a]}
		Robert Berger,\textsuperscript{[c]}
		Hendrike Braun,\textsuperscript{[a]} 
		Arno Ehresmann,\textsuperscript{[a]}
            Guido W. Fuchs,\textsuperscript{[a]}
		Thomas F. Giesen,\textsuperscript{[a]}
            Jochen Mikosch, \textsuperscript{[b]}
		Rudolf Pietschnig,\textsuperscript{[d]} 
	    and Arne Senftleben*\textsuperscript{[a]}
	\end{minipage}
}

\newcommand{\affiliation}{
\begin{itemize}

\item[{[a]}] Sudheendran Vasudevan, Simon T. Ranecky, Lutz Marder, Catmarna K\"ustner-Wetekam, Nicolas Ladda, Sagnik Das, Tonio Rosen, Vidana Popkova, Dr. Hangyeol Lee, Dr. Andreas Hans, Prof. Dr. Thomas Baumert, Dr. Hendrike Braun, Prof. Dr. Arno Ehresmann, Dr. Guido W. Fuchs, Prof. Dr. Thomas F. Giesen,  Dr. Arne Senftleben*\\
Institut f\"ur Physik and CINSaT, Universit\"at Kassel, Heinrich-Plett-Str. 40, 34132 Kassel, Germany\\
E-mail: arne.senftleben@uni-kassel.de

\item[{[b]}] Prof. Dr. Jochen Mikosch \\
Institut f\"ur Physik, Universit\"at Kassel, Heinrich-Plett-Str. 40, 34132 Kassel, Germany\\

\item[{[c]}] Dr. Steffen M. Giesen, Dr. Manjinder Kour, Prof. Dr. Robert Berger
Fachbereich Chemie, Philipps Universit\"at Marburg, Hans-Meerwein Str.4, 35032 Marburg, Germany\\

\item[{[d]}] Igor Vidanovi\'{c}, Dr. Denis Kargin, Prof. Dr. Rudolf Pietschnig\\
Institut f\"ur Chemie and CINSaT, Universit\"at Kassel, Heinrich-Plett-Str. 40, 34132 Kassel, Germany\\

\item[{[e]}] Dr. Tim Sch\"afer \\
Institut f\"ur  Physikalische Chemie, Georg-August-Universit\"at G\"ottingen, Tammannstr. 6, 37077 G\"ottingen, Germany\\

\end{itemize}
}

\renewcommand{\abstract}{Excited electronic states of fenchone, thiofenchone, and selenofenchone are characterized and assigned with different gas-phase spectroscopic methods and \textit{ab initio} quantum chemical calculations. With an increasing atomic number of the chalcogen, we observe increasing bathochromic (red) shifts, which vary in strength for Rydberg states, valence-excited states, and ionization energies. The spectroscopic insight is used to state-resolve the contributions in multi-photon photoelectron circular dichroism with femtosecond laser pulses. This is shown to be a sensitive observable of molecular chirality in all studied chalcogenofenchones. Our work contributes new spectroscopic information, particularly on thiofenchone and selenofenchone. It may open a perspective for future coherent control experiments exploiting resonances in the visible and or near-ultraviolet spectral regions.
}

\newcommand{\keywords}{
    Chirality \textbullet\ 
	Light-matter interactions \textbullet\ 
	Photoelectron spectroscopy \textbullet\ 
	Femtochemistry \textbullet\ 
	Circular dichroism
}

\begin{document}

\twocolumn[\vspace{-1.5cm}\maketitle\vspace{-1cm}
	\textit{\dedication}\vspace{0.4cm}]
\small{\begin{shaded}
		\noindent\abstract
	\end{shaded}
}

\begin{figure} [!b]
\begin{minipage}[t]{\columnwidth}{\rule{\columnwidth}{1pt}\footnotesize{\textsf{\affiliation}}}\end{minipage}
\end{figure}


\section{Introduction}
\label{introduction}

Sensing the chirality of molecules in the gas phase has become a popular scientific frontier. Taking advantage of the interaction-free environment coherent microwave spectroscopy,\cite{Patterson.2013a, Patterson.2013b, Shubert.2014} Coulomb explosion imaging,\cite{Pitzer.2013, Herwig.2013,  Pitzer.2016a, Pitzer.2016b} and circular dichroism (CD) using high harmonic spectroscopy \cite{Bayku.2018} or ion-yield \cite{Boesl.2006,Li.2006,Breunig.2009,Horsch.2011,Boesl.2013, Boesl.2016, Ring.2021} have been demonstrated in the last decade. Moreover, photoelectron circular dichroism (PECD) has been established as a versatile probe of chirality.\cite{Sparling.2025} PECD is a forward/backward asymmetry in the angular distribution of electrons derived from randomly oriented chiral molecules by ionization with circularly polarized light. It results from the scattering of the departing photoelectron with the potential created by the atomic cores of the molecule. The asymmetry in the photoelectron angular distribution changes sign either by changing the enantiomer or the helicity of the circularly polarized light. The strength of the asymmetry can be measured by the  Linear PECD (LPECD) quantity.\cite{Lehmann.2013, Lux.2015b} Since PECD is attributed purely to electric dipole interaction,\cite{Ritchie.1976} it exhibits a stronger effect than other chiroptic effects based on electric quadrupole or magnetic dipole interaction such as CD.\cite{Boesl.2013, Boesl.2016, Andrews.2020} 

After the theoretical prediction by Ritchie in 1976,\cite{Ritchie.1976} it took more than 25 years before the first experimental observation of PECD was reported using synchrotron radiation.\cite{Bowering.2001} For ionization with XUV and X-ray photons, the scattering electron that probes the ionic potential can originate either from core shells\cite{Hergenhahn.2004, Nahon.2006, Hartmann.2019, Catone.2022} or valence shells\cite{Stranges.2005, Garcia.2003c, Turchini.2004, Garcia.2008, Powis.2008, Nahon.2015, Catone.2022} of the molecule, depending on the photon energy. 

Pioneering experiments in the optical domain, reported a decade later, showed that the PECD effect can be equally strong using resonance-enhanced multi-photon ionization\-(REMPI) with femtosecond \cite{Lux.2012, Lehmann.2013} and nanosecond \cite{Kastner.2019} lasers. The principle of REMPI for the example of a $2+1$ scheme is sketched in Figure~\ref{Exp_setup}~(B): Where two photons are necessary to reach a bound excited state as an intermediate resonance, one additional photon ionizes the molecules. The existence of the intermediate resonances increases the multi-photon ionization rate drastically, making REMPI well-known for its high selectivity and sensitivity. \cite{Boesl.2021} REMPI-PECD experiments hence raise the obvious question, in which way excited states of the chiral molecule contribute in detail to the observed PECD effect. This question continues to be of high interest, despite and because of the increasing number of studies utilizing REMPI-PECD. \cite{Lux.2012, Lehmann.2013, Lux.2015b, Kastner.2017, Ranecky.2022, Beaulieu.2016, Rafiee.2014, Rafiee.2016, Beaulieu.2018} 

Fenchone has become a benchmark molecule in experimental and theoretical studies, turning it into the ``hydrogen atom of gas-phase chirality.''\cite{Kastner.2019, Kastner.2020, Kastner.2017, Beaulieu.2016b, Comby.2016, Blanchet.2021, Muller.2018,  Muller.2020, Kutscher.2023} It serves as a reference for enantiomeric excess determination with different methods. \cite{Kastner.2016, Comby.2023} In addition, many experimental and theoretical studies to understand its spectroscopic properties \cite{Pulm.1997, Devlin.1997, Goetz.2017, Singh.2020, Powis.2023} and internal dynamics \cite{Comby.2016, Blanchet.2021, Lee.2022} have been reported. Single-photon excitation of fenchone requires wavelengths in the ultraviolet. Derivates in which the oxygen atom is replaced with heavier chalcogens exhibit bathochromic (red) shifts in the absorption spectrum, which might make them more accessible for future coherent control experiments with visible wavelengths. 

Here, we investigate the excited states and the linear REMPI-PECD for fenchone and its heavier derivates, thiof\-enchone (oxygen substituted by sulfur) and selenofenchone (oxygen substituted by selenium). In the terminology of fenchone as the ``hydrogen atom of gas-phase chirality,'' the other chalcogenofenchones are its alkali atoms. While both enantiomers of fenchone are readily commercially available, thiofenchones and selenofenchones have to be synthesized. Consequently, much less spectroscopic information on Rydberg states, valence-excited states, and ionization energies is available as compared to fenchone. To the best of our knowledge, REMPI photoelectron spectroscopy has not been reported for thiofenchone nor any gas phase spectroscopy on selenofenchone. For thiofenchone, the ionization energy has been reported,\cite{Guimon.1974} and the $n=4$ Rydberg state and a $\pi\rightarrow\pi^*$ state have been identified with gas phase UV absorption spectroscopy.\cite{Falk.1988} For selenofenchone, circular dichroism and UV-VIS absorption spectra in various solvents have been published, identifying a $n\rightarrow\pi^*$ excitation in the visible spectral region.\cite{Wijekoon.1983, Andersen.1982} These solution-phase studies have compared the UV-VIS absorption bands of selenofenchone with those of thiofenchone and fenchone, observing a bathochromic shift with increasing atomic number of the chalcogen. 

This manuscript is organized as follows. First, we describe our methodology. Experimentally, we use single-phot\-on VUV absorption spectroscopy, nanosecond laser REMPI spectroscopy, and wavelength scanning multi-photon photoelectron spectroscopy using picosecond duration laser pulses. In addition, we employ angle-resolved photoelectron spectroscopy to measure REMPI-PECD using femtosecond laser pulses. As a theoretical approaches, we use quantum chemical calculations on the density functional theory (DFT) and coupled cluster (CC) level. Synthesis and sample characterization of thiofenchone and selenofenchone are detailed in an accompanying paper. For fenchone, thiofenchone, and selenofenchone, we use our combined results from experiment and theory to obtain the adiabatic \textit{{I\textsubscript{P}}} and assign excited electronic states. Assisted by calculated values for tellurofenchone and polonofenchone, we will discuss the scaling of excited state energies in chalcogenofenchones. In a subsequent section, we discuss femtosecond REMPI-PECD experiments on fenchone, thiofenchone and selenofenchone at a fixed wavelength of \SI{376}{\nano\metre} and use the obtained spectroscopic insight to resolve the contributions of individual states.


\section{Experimental techniques}
\label{Exp_tech}

\subsection{Sample preparation}
The chalcogenofenchones have two stereocenters, but due two the bicyclic structure only two enantiomers are geometrically possible: $(1R, 4S)$ and $(1S, 4R)$. We will refer to the $(1R, 4S)$-enantiomers as $(R)$-chalcogenofenchone and to the $(1S, 4R)$ as $(S)$ while also giving the sign of the optical rotation. Commercially available (\textit{S})-($+$)-fenchone (Acros, purity $\ge\SI{97}{\percent}$, enantiomeric excess (e.e.) $\approx\SI{99}{\percent}$) and (\textit{R})-($-$)-fenc\-hone (Merck, purity $\ge\SI{98}{\percent}$, e.e. $\approx\SI{84}{\percent}$) were used without further processing. The synthesis and characterization of thiofenchone and selenofenchone will be described in detail in an accompanying paper and briefly summarized here. (\textit{S})-($+$)- and (\textit{R})-($-$)-thiofenchones were synthesized from the respective fenchone enantiomers with Lawesson's reagent (Alfa Aesar) in an \textit{o}-xylene solution at \SI{155}{\celsius}.~(\textit{S})-($+$)- and (\textit{R})-($-$)- selenofenchone were synthesized from the respective fenchone enantiomers suspended in purified mesitylene using bis(1,5-cyclooctanediylboryl) mo\-noselenide at \SI{120}{\celsius}. The latter was prepared \textit{in situ} from 9-borabicyclo[3.3.1]nonane-dimer (Merck) at elevated temperatures above \SI{150}{\celsius} according to literature procedur\-es.\cite{Wrackmeyer.2018,Shimada.1997}

\subsection{Nanosecond \texorpdfstring{$2+1$ R} REMPI spectroscopy}
 
Linearly polarized \unit{\nano\second} laser pulses were intersected with a pulsed molecular beam in the interaction region of a velocity-map imaging (VMI) spectrometer\cite{Eppink.1997}, which was operated as a time-of-flight mass spectrometer. Here, we integrated all detected masses to obtain the total ion yield. The cold molecular beam was created by co-expanding helium and the respective molecular sample, which was heated to about \SI{70}{\celsius}.\cite{Westphal.2019, Westphal.2020} For fenchone, we used a frequency-doubled commercial dye laser to produce wavelengths between \num{412} and \SI{417}{\nano\metre} (\textit{LIOP-TEC}, pumped by the second harmonic of a Nd:YAG laser at \SI{532}{\nano\metre}, dye: Styryl 9). For thiofenchone and selenofenchone, we used the fundamental of a different commercial dye laser (\textit{Sirah}, pumped by the third harmonic of a Nd:YAG laser at \SI{355}{\nano\metre}) to produce wavelengths from \SIrange{441}{446}{\nano\metre} (dye: Coumarin 120) and from \SIrange{463}{468}{\nano\metre} (dye: Coumarin 47) without frequency doubling. The laser pulse energy was in the range of \SIrange{0.5}{1}{\milli\joule} and the bandwidth below \SI{.2}{\nano\meter}. A plano-convex lens of \SI{300}{\milli\metre} focal length was used to focus the pulses in the interaction region of the VMI spectrometer. The repetition rate of the laser was \SI{10}{\hertz}.

The $3p$ Rydberg states of fenchone (see Figure~\ref{Ns_REMPI_Fen}~(B)) were characterized with a different experimental setup. A frequen\-cy-doubled commercial dye laser was used to produce laser wavelengths between \num{383} and \SI{395}{\nano\metre} (\textit{Sirah}, \textit{Co\-bra-Stretch}, pumped by the second harmonic of a Nd:YAG laser at\- \SI{532}{\nano\metre}, dye: Styryl 11). The laser pulse energy was around \SI{1.1}{\milli\joule} at a repetition rate of \SI{20}{\hertz} and a bandwith below \SI{.5}{\nano\metre}. The pulses were linearly or circularly polarized and focused into a VMI spectrometer operated in time-of-flight mode, similar to the setup described above, but with a moderately cold continuous molecular beam of fenchone seeded in helium.

\subsection{Multi-photon photoelectron spectroscopy}
\label{ps_PES}

Starting from femtosecond pulses from a \SI{3}{\kilo\hertz} repetition rate titanium-sapphire amplifier, we generated $\approx\SI{0.4}{\pico\second}$ long pulses with a spectral width of $\approx\SI{.6}{\nano\metre}$ by second harmonic generation using a \SI{5}{\milli\metre} thick $\beta$-BBO crystal. By changing the phase-matching angle, we tuned the wavelength of the second harmonic from \SIrange{375}{411}{\nano\metre} (\SIrange{6}{9}{\micro\joule} pulse energy), as characterized with a commercial grating spectrometer (\textit{Avantes ULS3648}). A plano-convex lens with \SI{250}{\milli\metre} focal length was used to focus the pulses in the interaction region of a VMI spectrometer \cite{Eppink.1997}. The molecular sample was introduced by backfilling the vacuum chamber to a pressure of about \SI{4e-6}{\milli\bar}. From the recorded two-dimensional photoelectron images, the cylindrically symmetric three-dimensional photoelectron momentum distributions were reconstructed by a rBasex algorithm implemented in PyAbel.\cite{Hickstein.2019} From the momenta, the photoelectron energies were calculated. The energy axis was calibrated with photoelectrons derived from ionizing xenon atoms with the third harmonic of a Q-switched Nd:YAG ns laser at \SI{355}{\nano\metre}.

\subsection{Single-photon VUV absorption}
\label{sec:VUV_absorption}
A home-built absorption set-up has been used to measure single-photon gas phase VUV absorption spectra. A deuterium lamp (\textit{Hamamatsu L11798}) served as light source, which produced a continuous -- though strongly varying in intensity, leading to a lower signal-to-noise ratio at higher wavelengths -- spectrum from \SIrange{115}{400}{\nano\metre}. The spectrum has been dispersed by a commercial normal-incidence grating spectrometer with \SI{1}{\metre} focal length in Rowland geometry (\textit{McPherson Type 225} with 1200 lines/mm).\cite{Mortensen.2017,Yen.1981} The dispersed light was focused on a photodiode behind the exit slit. A motorized grating rotation/translation system was used to scan the desired spectral range. A \SI{2}{\centi\metre} path length target cell mounted between the deuterium lamp and the spectrometer's entrance slit was equipped with magnesium fluoride windows. Liquid samples were filled into the target cell at room temperature, which was briefly evacuated to remove residual gas before being heated up to \SI{60}{\celsius} for fenchone and to \SI{40}{\celsius} for thiofenchone and selenofenchone. A reference measurement of the lamp spectrum without sample was used to calculate the relative absorption via the Beer-Lambert law. The required high vapor pressure of the samples inside the cell led to the rapid formation of deposits on the windows and, consequently, short time frames for each measurement. Thus, 15 spectra were measured for each molecule, averaged and smoothed in segments, and stitched together. The resulting spectral resolution in the experiments was estimated to be about \SI{0.1}{\electronvolt}. The wavelength axis has been calibrated using the absorption spectrum of nitrogen\cite{Babcock.1976} and the emission spectrum of a mercury arc lamp.

\subsection{Femtosecond \texorpdfstring{$2+1$ R} REMPI-PECD}

\begin{figure*}
\begin{center}
\includegraphics[width=17.2cm]{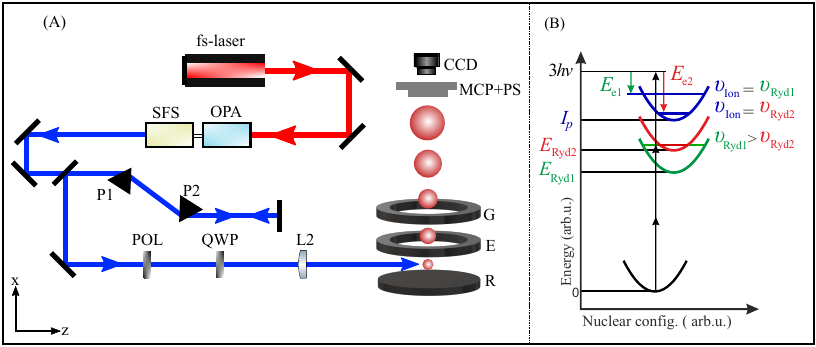}
\caption{(A) Experimental setup for femtosecond REMPI-PECD. It consists of a titanium-sapphire femtosecond amplifier, an optical parametric amplifier with sum-frequency conversion stage (OPA/SFS), a prism compressor (P1, P2) for pulse compression, a nanowire grid polarizer (POL) and a quarter-waveplate (QWP) to create circularly polarized light, a UV fused silica plano-convex lens (L2), the VMI electrodes repeller (R), extractor (E), and ground (G), a microchannel plate detector (MCP), phosphor screen (PS) and camera (CCD). In the VMI apparatus, the laser pulses propagate along the $z$-axis, while the particles are accelerated along the $x$-axis. (B) Schematic of the $2+1$ REMPI processes involving two electronic Rydberg states at energies $E_\text{Ryd1}$ and $E_\text{Ryd2}$. First, two-photon excitation populates these states at different vibrational states $v_\text{Ryd1} > v_\text{Ryd2}$. Due to the Rydberg nature, the following ionization by another photon hardly changes the vibrational excitation, which gives rise to higher photoelectron energy $E_{e2}$ from the higher-lying $E_\text{Ryd2}$ Rydberg state.}
\label{Exp_setup}
\end{center}
\end{figure*}

The optical layout of the femtosecond PECD experiment is shown in Figure \ref{Exp_setup} (A). Laser pulses from a \SI{3}{\kilo\hertz} repetition rate titanium-sapphire amplifier centered at \SI{785}{\nano\metre} (\textit {Femtopower HE}) with a pulse duration of below \SI{25}{\femto\second} and a pulse energy of \SI{0.4}{\milli\joule} were used to pump a commercial optical parametric amplifier (OPA) with a subsequent frequency conversion stage (\textit{LightConversion TOPAS prime} with \textit{NirUVis} extension), which produced output pulses centered at \SI{376}{\nano\metre} with a spectral width of \SI{8.3}{\nano\metre}. A thin nano-wire grid broadband polarizer (\textit{Quantum Design} \SIrange{300}{1000}{\nano\metre}) was used to ensure that the polarization of linearly polarized light was perpendicular to the acceleration direction of the spectrometer (see below and Figure~\ref{Exp_setup} (A)). Circularly polarized light was generated by an achromatic quarter waveplate (\textit{B.Halle}, \SIrange{300}{470}{\nano\metre}). The circularity of the light was in all cases well above 99\%, as measured via the Stokes parameter $\vert$S3$\vert$.\cite{Mcmaster.1954} A plano-convex fused silica lens with \SI{250}{\milli\metre} focal length (\textit{Eksma Optics}) was used to focus the UV laser pulses with \SI{3}{\micro\joule} energy into the interaction region of a VMI spectrometer \cite{Eppink.1997}, which can be operated for ions or electrons. 

A prism compressor consisting of two UV fused silica prisms was used to compensate for chirp, and bandwidth-limited pulses were achieved in the interaction region by maximizing the multi-photon ionization signal of xenon. Assuming transform-limited compression, the pulse duration was estimated to be around \SI{25}{\femto\second} (FWHM). For the measured Gaussian beam spot radius of \SI{48}{\micro\metre}, we calculated a peak intensity of \SI{3e12}{\watt\per\centi\metre\squared}. The Keldysh parameter \cite{Keldysh.1998} $\gamma$ was estimated to be around $10$, confirming the multi-photon ionization regime. The sample supply lines between the reservoir and the nozzle were heated to about \SI{60}{\celsius}. The nozzle (\textit{Agarscientific}, diameter: \SI{100}{\micro\meter}) was heated to about \SI{90}{\celsius}. The sample reservoir outside the vacuum chamber was heated to about \SI{30}{\celsius} for fenchone, while for thiofenchone and selenofenchone it was heated to about \SI{50}{\celsius}. The sample was introduced by backfilling the vacuum chamber to a pressure of about \SI{3e-6}{\milli\bar}.  Before the PECD experiment, we measured the mass spectra to ensure we had only the desired molecular substance (see Appendix).

Photoionization occurred between the repeller and extractor electrodes of the VMI spectrometer (see Figure \ref{Exp_setup} (A)). Photoelectron momentum distributions were recorded on a position-sensitive detector, a microchannel plate (MCP) coupled to a phosphor screen imaged by a CCD camera (\textit{Lumenera Lw165m}). We accumulated events over \num{975000} laser shots (\SI{130}{\milli\second} exposure time, 2500 images) for each of the left-circular (LCP), right-circular (RCP), and linear (LIN) polarization. After every 500 images, the polarization was switched to minimize artifacts due to long-term experimental drifts. PECD images were derived by subtracting RCP from LCP photoelectron angular distributions. The resulting two-dimensional distributions were Abel-inverted to reconstruct the three-dimensional momentum distributions using the rBasex algorithm implemented in PyAbel.\cite{Hickstein.2019} The Abel inversion is performed using Legendre expansions of the electron momentum distributions. The retrieved Legendre coefficients $c_i$ allow us to calculate the linear photoelectron circular dichroism (LPECD): \cite{Lux.2015b}
\begin{equation}
 \label{LPECD}
    \textrm{LPECD} = \frac{1}{c_0}\left( 2c_1-\frac{1}{2}c_3 +\frac{1}{4}c_5 \right).
\end{equation}
To account for the different enantiomeric excesses of the samples, LPECD values for (\textit{R})-($-$)-fenchone were multiplied by the factor 1/0.84. We proceeded likewise for the (\textit{R}) enantiomers of thiofenchone and selenofenchone to check if the enantiomeric excess is conserved during synthesis.

\section{Computational methodology}

All the calculations presented in this article were carried out with the TURBOMOLE (V7.6.0) program package.\cite{TurboMole} Electronic ground-state structures for all chalcogenof\-encho\-nes were optimized for the energy gradient using density-func\-tio\-nal theory (DFT) with the long-range corrected hybrid functional CAM-B3LYP.\cite{Yanai.2004, Tawada.2004} The resulting structures were confirmed to be minima on the respective potential energy surfaces by a harmonic vibrational frequency analysis.
In all systems, the aug-cc-pVTZ basis set was chosen for C and H. For an improved description of the Rydberg states, quadruply augmented basis sets were used for the chalcogen centers. Specifically, we employed q-aug-cc-pVTZ for O, q-aug-cc-pV(T+d)Z for S, and q-aug-cc-pVTZ-PP for Se, Te, and Po as the basis sets. Additionally, the calculations were repeated with diffuse functions of these basis sets anchored to the center of mass of the systems, which was confirmed not to impact the results to any significant degree.
Relativistic small-core pseudopotentials were used on Se, Te, and Po. 

All calculations applied the multipole-accelerated resolut\-ion\--of\--identity (MARI-J) approximation for the Coulomb energy contribution using the corresponding RI aug\--cc\--pV\-TZ basis sets (jbas) for all the atoms. The m4 integration grid was used for numerical integration to determine exchange--correlation contributions. The threshold for the self-consistent field (SCF) energy convergence was set at $10^{-9} E_\mathrm{h}$. Structure optimizations were performed until the norm of the
energy gradient fell below $5\times 10^{-4} E_\mathrm{h} a_0^{-1}$ and the displacements were smaller than $5\times 10^{-4} a_0$. The norm of the analytic gradients of the energy with respect to displacements of the nuclei was below $5\times 10^{-3} E_\mathrm{h} a_0^{-1}$, and the estimated energy change between two optimization steps was below $5\times 10^{-6} E_\mathrm{h}$. Single-point calculations for each of the cations were performed at the same level of theory -- using the equilibrium structure of the neutral electronic ground state -- to obtain the vertical ionization energies.

The ground-state energy minimum structure was used to calculate vertical excitation energies
and oscillator strengths
using time-dependent DFT (TDDFT), as implemented in the Turbomole package.
At least 20 low-lying electronically excited states were calculated for each system, according to the experimental energy ranges.
Furthermore, two-photon absorption (TPA) cross sections for linear and circular polarized light were obtained at default settings using TDDFT.

To evaluate the significance of the TDDFT results, we compare them to the wave function-based second-order approximate coupled cluster method CC2,\cite{Christiansen.1995} using the implementation in Turbomole. CC2 is expected to offer energies at the quality of second-order Møller-Plesset perturbation theory (MP2) quality while giving access to excited state energies and transition moments.  The excited-state energies converged to an accuracy of $10^{-7} E\textsubscript{h}$, and the vector function converged to $10^{-4} E\textsubscript{h}$ while employing a resolution of the identity parametrization for both the Coulomb and exchange integrals (RI-JK).

To assign the \(m\)-quantum number related character of $p$ ($p_x$, $p_y$, $p_z$) and $d$ ($d_{xy}$, $d_{xz}$, $d_{yz}$, $d_{z^2}$, $d_{x^2-y^2}$) and higher \(l\)-quantum number
Rydberg orbitals, we computed natural transition orbitals (NTOs) and visualized them for assignment.
NTOs are obtained by using the singular vectors from a singular value decomposition of the transition density \(P^{ia}\) between states \(i\) and \(a\) to transform the orbitals of the system.
\begin{equation}
    P^{ia}=UDV^\dagger,    
\end{equation}
where \(D\) is a diagonal matrix containing the singular values that determine the contribution
of each NTO for the transition.
The left singular vectors \(U\) are then related to the \textit{hole} left by the excited electron from the initial state \(i\), corresponding approximately to an occupied orbital determined by
the orbital mixing coefficients \(U\). Equivalently, the right singular vectors \(V\) describe the final state \(a\) of the excited \textit{particle}, corresponding approximately to unoccupied molecular orbitals of valence- or Rydberg-type.

\begin{figure*}
    \centering
    \includegraphics[width=17.2cm]{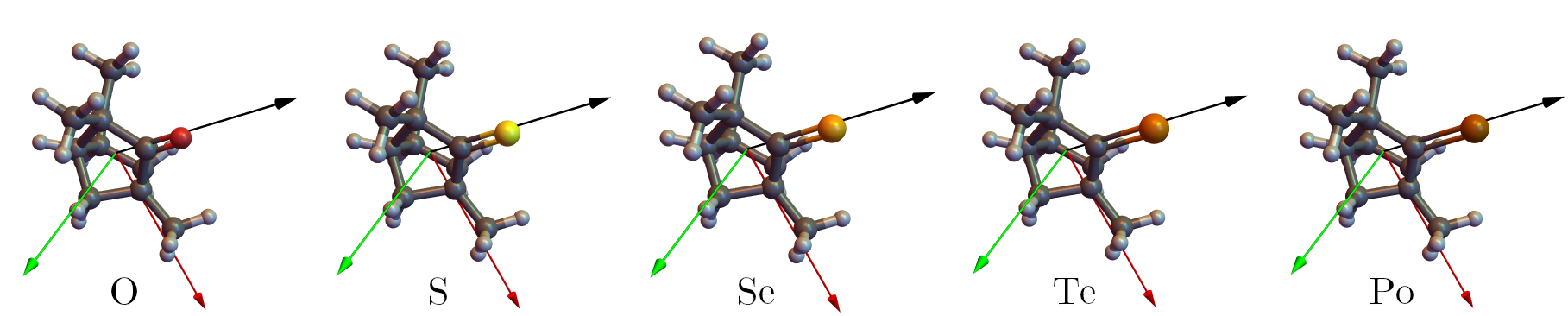}
    \caption{Optimized molecular structures of the ($1R$, $4S$)-chalcogenofenchones as identified by the atomic symbols of the chalcogenes. The green, red, and black arrows show the chosen
    \(x\), \(y\), and \(z\)-axes, respectively.}
    \label{fig:moleculestructures}
\end{figure*}

A coordinate system based on the local symmetry of the formaldehyde-like group containing the chalcogen atom was introduced to identify the $m$-quantum number related character of the orbitals.
Choosing the C=X bond (with X either O, S, Se, Te or Po) as the \(z\)-axis, the plane encompassing the next-neighboring C atoms
is chosen to contain the $y$-axis, with the $x$-axis normal to this plane (cf. Figure \ref{fig:moleculestructures}).
It should be noted that the Rydberg-like NTOs are not always aligned with this coordinate system
perfectly and may also choose another axis than $z$ as their principal axis.
We chose to keep the coordinate system fixed for all molecules and states, and instead change
the basis according to the shape and orientation of the NTOs. That means for example,
that if a \(d_{zz}\)-like orbital is rotated so that its principal axis is the \(x\)-axis
in our choice of coordinates, we denote it as \(d_{xx}\), with the other orbitals for the same
\(l\)-quantum number also adjusted to the new principal axis, e.g. \(d_{xx-yy}\rightarrow d_{yy-zz}\), so that the choice is consistent.

\section{Assignment of excited states through spectroscopy}
\label{sec:spectrocopy}

In this section, we present our findings of the spectroscopic experiments on fenchone, thiofenchone, and selenofenchone. For each molecule, the discussion is organized in the following way: As a first step, we assign the lowest lying Rydberg state of $s$ symmetry using \unit{\nano\second} $2+1$ REMPI spectra. With this, we determine the adiabatic ionization energy $I_P$ using, in addition, multi-photon photoelectron spectra at different excitation wavelengths ranging from \SIrange{375}{411}{\nano\metre}. Then, we assign further electronic states using the adiabatic $I_P$, multi-photon photoelectron spectra, and single-photon VUV absorption spectra together with \textit{ab initio} quantum chemical calculation.

\subsection{Fenchone}
\label{sec:fenchone}

\begin{figure*}
	\begin{center}
		\includegraphics[width=17.4cm]{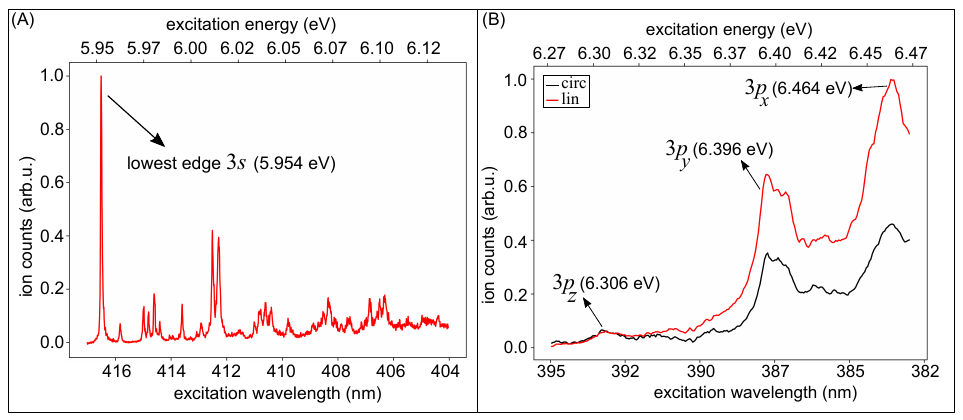}
		\caption{Nanosecond $2+1$ REMPI spectrum of fenchone covering (A) the lower vibrational levels of the $3s$ Rydberg state measured with linearly polarized laser pulses at high spectral resolution and (B) the $3p$ Rydberg states measured with linear (red) and circular (black) laser polarizations.}
		\label{Ns_REMPI_Fen}
	\end{center}
\end{figure*}

Figure \ref{Ns_REMPI_Fen}~(A) shows a highly-resolved \unit{\nano\second} $2+1$ REMPI spectrum of fenchone. The intense peak at \SI{416.5}{\nano\metre} is the onset of the spectrum and indicates the threshold for reaching the $3s$ Rydberg state. Considering the two-photon character for reaching this resonance, we obtain \SI{5.954(2)}{\electronvolt} as the energy difference between the $3s$ Rydberg state and the electronic ground state, which agrees with earlier experiments.\cite{Kastner.2020, Singh.2020, Powis.2023} At \SI{6.300}{\eV}, the TDDFT model overestimates the measured value, while CC2 underestimates it at \SI{5.645}{\eV}. As the calculations often yield more precise relative energies between different excited states than the absolute energies, Tables~\ref{fen_TDDFT} and~\ref{fen_CC2} also contain columns of energies $E^*_\text{calc}$ that are shifted by the difference of the respective calculated $3s$ energy to the measured value. In addition to the intense peak attributed to the vibrational ground state of the $3s$, the spectrum displays a pronounced progression - the partially resolved vibrational structure of the $3s$ electronic state. 

Figure \ref{Ns_REMPI_Fen}~(B) displays a \unit{\nano\second} $2+1$ REMPI spectrum of the $3p$ Rydberg states recorded with linearly and circularly polarized light. From this measurement, we obtained \SI{6.306(7)}{\electronvolt} for the weak signal of the lowest-energy $3p$ state. Following the suggestion of Powis \etal\cite{Powis.2023} we assign this state as $3p_z$. It has a \SI{.352(7)}{\eV} higher energy than the $3s$ state. This difference is nearly matched by the CC2 calculation at \SI{.352}{\electronvolt} but overestimated by the TDDFT model at \SI{.438}{\electronvolt}. The $3p_z$ state assignment is supported by the observation of a relatively high signal for circular polarization as compared to linear polarization, in agreement with existing calculations\cite{Powis.2023, Singh.2020} and with our calculated two-photon cross sections in Table~\ref{fen_TDDFT}. We must add that comparing the $2+1$ REMPI signal strength with calculated two-photon cross sections is only reasonable if the ionization step is saturated, i.e., the excited molecules are likely to be ionized. This has been verified at similar laser conditions by Singh \etal\cite{Singh.2020}

\begin{table*}[htbp]
    \caption{\label{fen_TDDFT} TDDFT results for fenchone: Vertical electronic singlet excitation energies $E_\text{calc}$, calculated energies $E^*_\text{calc}$ shifted such that the $3s$ energy matches the measured value, experimental state energies $E_\text{exp}$, two-photon excitation cross sections for linearly (circularly) polarized light $\sigma_\text{lin}$ ($\sigma_\text{circ}$) and oscillator strengths $f$. Parentheses mark a tentative assignment.}
    \begin{tabular}{l S[round-mode=places,round-precision=3,group-digits = none] S[round-mode=places,round-precision=3,group-digits = none] S S[round-mode=places,round-precision=4,group-digits = none] S[round-mode=places,round-precision=4,group-digits = none] S[round-mode=places,round-precision=4,group-digits = none]}
    \toprule
    {Electronic transition} & {\centering{$E_\text{calc}$ (\unit{\electronvolt})}} & \centering{{$E^*_\text{calc}$ (\unit{\electronvolt})}} & \centering{{$E_\text{exp}$ (\unit{\electronvolt})}} & {$\sigma_\text{lin}$ ($10^{-50}\unit{\centi\metre\tothe{4}\second}$)} & {$\sigma_\text{circ}$ ($10^{-50}\unit{\centi\metre\tothe{4}\second}$)} & {$f$} \\
    \midrule
    $n_y\rightarrow \pi^*_x$ & 4.336&3.990& & 0.00000397& 0.00000398& 0.00001673\\
    $n_y\rightarrow 3s$ &6.300&5.954& 5.954& 0.01084673& 0.01609737& 0.00040326\\
    $n_y\rightarrow 3p_z$ & 6.738&6.392& 6.306 &  0.00316360 & 0.00470488 & 0.02490511\\
    $n_y\rightarrow 3p_y$ & 6.770&6.424&6.396& 0.03098990 & 0.01243724 & 0.01513925\\
    $n_y\rightarrow 3p_x$ & 6.816&6.4700&6.464  & 0.01124630 & 0.00403201 & 0.00404011\\
    $n_y\rightarrow 3d_{xx}$ & 7.226&6.880& 6.85 & 0.01203027 & 0.01653848 & 0.00544779\\
    $n_y\rightarrow 3d_{yz}$ & 7.338&6.992&  & 0.05526848 & 0.00740340 & 0.00750651\\
    $n_y\rightarrow 3d_{xy}$ & 7.351&7.005&  & 0.01854577 & 0.01219580 & 0.00056347\\
    $n_y\rightarrow 3d_{yy-zz}$ & 7.380&7.034& {\multirow{2}{*}{7.04}} & 0.01079528 & 0.00443590 & 0.01546475\\
    $n_y\rightarrow 3d_{xz}$ & 7.409&7.063&  & 0.00704345 & 0.00689820 & 0.00103540\\
    $n_y\rightarrow 4s$ & 7.454&7.108&  & 0.00206445 & 0.00166777 & 0.00070828\\ 
    $\sigma_{1}\rightarrow \pi^*_x$ & 7.578&7.232 &  & 0.00191944 & 0.00081601 & 0.01729118\\ 
    $n_y\rightarrow 4p_x$ & 7.601& 7.255 &  & 0.00028075 & 0.00018493 & 0.00055264\\
    $n_y\rightarrow 4p_z$ & 7.613&7.267 &  & 0.00011112 & 0.00006147 & 0.00422059\\
    $n_y\rightarrow 4p_y$ & 7.620&7.274 &  & 0.00046744 & 0.00019486 & 0.00180499\\ 
    $n_y\rightarrow 5p_x$ & 7.855&7.509& {\multirow{2}{*}{7.46}}& 0.00133413 & 0.00189606 & 0.00022711\\
    $n_y\rightarrow 5p_y+5p_{(z)}$ & 7.878&7.532 &  & 0.00422934 & 0.00275318 & 0.00536995\\
    $n_y\rightarrow 5p_z-5p_{(y)}$ & 7.912&7.566 &  & 0.00065686 & 0.00043630 & 0.00168727\\
    $n_y\rightarrow 4d_{(yz)}$ & 7.917&7.571 &  & 0.00288016 & 0.00425967 & 0.01713826\\
    $n_y\rightarrow 4d_{(xx-zz)}$ & 7.934&7.588 &  & 0.00099353 & 0.00140398 & 0.00020223\\
    $n_y\rightarrow 4d_{(yy)}$ & 7.952&7.606 &  & 0.00281157 & 0.00321463 & 0.00430929\\
    $n_y\rightarrow 4d_{(xy)}$ & 8.018&7.672&   & 0.00369505 & 0.00402508 & 0.00612442\\
    $n_y\rightarrow 4d_{(xz)}$ & 8.027&7.681&   & 0.00198603 & 0.00179581 & 0.00177733\\
    $\sigma_{-1}\rightarrow 3s$ & 8.086&7.74 &   & 0.01519057 & 0.00951063 & 0.00354163\\
    $n_y\rightarrow 4f_{(yyy-3xxy)}$ & 8.116&7.77 &  & 0.01229744 & 0.01742408 & 0.00602374\\
    $n_y\rightarrow 4f_{xxy}(+6s)$ & 8.160&7.814&   & 0.02307298 & 0.02122190 & 0.00030092\\
    $\sigma_{-3}\rightarrow 5s$ & 8.176&7.83&  & 0.00090853 & 0.00088510 & 0.01377537\\
    \bottomrule
    \end{tabular}\\
\end{table*}

For the second $3p$ state ($3p_{y}$), we measured an energy of \SI{6.396(7)}{\electronvolt}, which agrees with the earlier work of Powis.\cite{Powis.2023} The difference to the $3s$ energy is \SI{.442}{\electronvolt}, which is underestimated by \SI{.06}{\electronvolt} in the CC2 model and overestimated by \SI{.03}{\electronvolt} in the TDDFT. We suggest to assign the intense peak above \SI{384}{\nano\metre} in the \unit{\nano\second} $2+1$ REMPI spectrum (Figure~\ref{Ns_REMPI_Fen}~(B)) to the third $3p$ state ($3p_{x}$) with an energy of \SI{6.464(7)}{\electronvolt}. This suggestion differs from the recent work of Powis and Singh,\cite{Powis.2023} who locate the $3p_x$ band origin in the same broad peak as the $3p_y$ band origin. Our assignment is supported by the energy differences calculated by our TDDFT and CC2 models (See Tables~\ref{fen_TDDFT} and~\ref{fen_CC2}). Moreover, the observed lower circular-to-linear ratio of the $3p_x$ peak of ca.~0.46 compared to ca.~0.55 for the $3p_y$ peak fits well to our TDDFT results (see Table~\ref{fen_TDDFT}) and older calculations.\cite{Singh.2020, Goetz.2017} In energies relative to the $3s$ state, the calculated values for the $3p_y$ and $3p_x$ states are again slightly overestimated by the TDDFT model and more significantly underestimated by the CC2 model.

\begin{figure}[tbp]
	\begin{center}
		\includegraphics[width=8.4cm]{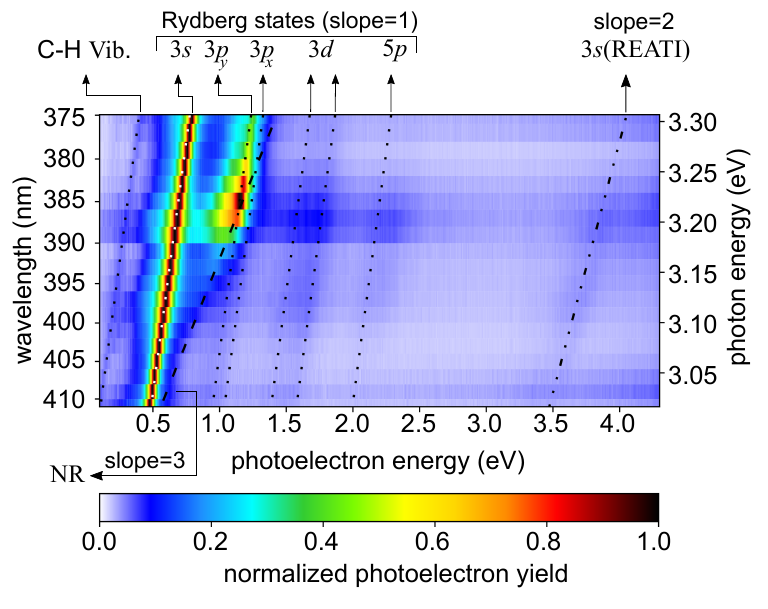}
		\caption{Multi-photon photoelectron spectra of fenchone. Each row in the figure represents a measurement at a different central wavelength. All measurements used a pulse duration of \SI{0.4}{\pico\second} and linear polarization. In each row, the signal is normalized to its maximum. Lines indicate linear scalings that can be attributed to different processes: $2+1$ resonance-enhanced multi-photon ionization through the labeled Rydberg states yields a slope of one, and $2+2$ resonance-enhanced above-threshold ionization (REATI) through Rydberg states a slope of two. In contrast, non-resonant (NR) multi-photon ionization is associated with a slope of three. The corresponding second-order Legendre coefficients are presented in Figure~\ref{Leg2_fen} in the appendix.}
		\label{psScan_FEN}
	\end{center}
\end{figure}

For further insight, multi-photon photoelectron spectra were measured for 19 different central wavelengths ranging from \SIrange{375}{411}{\nano\metre}. Figure~\ref{psScan_FEN} pre\-sents all photoelectron spectra together such that it constitutes a map of the photoelectron yield as a function of the photoelectron energy and the energy of the laser photons. In addition to the map of the yield, we also consider the map of the second-order Legendre coefficient of the photoelectron angular distributions (see Figure~\ref{Leg2_fen} in the appendix). These map-like representations simplify the identification of different ionization mechanisms due to linear scalings of the photoelectron energies $E_\text{kin}$ with the photon energy $h\nu$, represented by dotted lines in Figure~\ref{psScan_FEN}. Most ionization mechanisms are described very well by lines with a slope of one. This matches previous observations\cite{Kastner.2017, Singh.2021} for a $2+1$ REMPI process via a resonant Rydberg state (see Figure~\ref{Exp_setup}(B)). Due to the very similar potential energy surfaces of the neutral Rydberg states and the ionic ground state, the Franck--Condon factors for transitions with preserved vibrational quantum number dominate in the ionization step ($\Delta v=0$ propensity rule). \cite{Kastner.2017, Singh.2021} Under such conditions, the following equation for the photoelectron energy applies\cite{Kastner.2017}
\begin{equation} \label{eq:Kastner}
				E_\text{kin}(h\nu) = h\nu +E_\text{Ryd} - I_{P},   
\end{equation}
where $E\textsubscript{Ryd}$ denotes the Rydberg state energy and $I_{P}$ the ionization energy.

\begin{table*}[htbp]
    \caption{\label{fen_CC2} CC2 results for fenchone: Vertical electronic singlet excitation energies $E_\text{calc}$, calculated energies $E^*_\text{calc}$ shifted such that the $3s$ energy matches the measured value, experimental state energies $E_\text{exp}$, and oscillator strengths $f$. Parentheses mark a tentative assignment.}
    \begin{tabular}{c S[round-mode=places,round-precision=3,group-digits = none] S[round-mode=places,round-precision=3,group-digits = none] c S[round-mode=places,round-precision=4,group-digits = none]}
    \toprule
    \multicolumn{5}{c}{CC2 Fenchone}\\
    \midrule
    Electronic Transitions &\centering{{$E_\text{calc}$}(\unit{\electronvolt})}&\centering{{$E^*_\text{calc}$}(\unit{\electronvolt})} &\centering{{$E_\text{exp}$}(\unit{\electronvolt})}& \centering{{$f$}} \\
    \midrule
    $n_y\rightarrow \pi^*_x$ & 4.356 &4.665 &  & 0.00003\\
    $n_y\rightarrow 3s$ & 5.645 &5.954 &5.954  & 0.00315\\
    $n_y\rightarrow 3p_z$ & 5.980  &6.289 &6.306    & 0.01374\\
    $n_y\rightarrow 3p_y$ & 6.024    &6.333 & 6.396 & 0.01273\\
    $n_y\rightarrow 3p_x$ & 6.075    &6.384 & 6.464 & 0.00203\\
    $n_y\rightarrow 3d_{xx}$ & 6.483    &6.792  & 6.85 & 0.00582\\
    $n_y\rightarrow 3d_{xy}$ & 6.561    & 6.870 & & 0.00052\\
    $n_y\rightarrow 3d_{yz}$ & 6.565    & 6.874 &  & 0.00517\\
    $n_y\rightarrow 3d_{yy-zz}$ & 6.590    & 6.899 & {\multirow{2}{*}{7.04}}  & 0.01028\\
    $n_y\rightarrow 3d_{xz}$ & 6.609    &6.918&    & 0.00275\\
    $n_y\rightarrow 4s$ & 6.703  & 7.012& & 0.00085\\
    $n_y\rightarrow 4p_z$ & 6.882      & 7.191 &  & 0.00548\\
    $n_y\rightarrow 4p_x+4p_y$ & 6.890      &7.199 &    & 0.00296\\
    $n_y\rightarrow 4p_y+4p_x$ & 6.903      & 7.212&    & 0.00513\\
    $n_y\rightarrow 4f_{yzz}+4p_y$ & 7.084     &  7.393 &    & 0.00230\\
    $n_y\rightarrow 4d_{yz}$ & 7.116     &7.425 &    & 0.00739\\
    $n_y\rightarrow 4d_{xz}$ & 7.146     & 7.455 &    & 0.00114\\
    $n_y\rightarrow 4d_{(yy-zz)}$ & 7.178    & 7.487 &    & 0.00912\\
    $n_y\rightarrow 5p_x$ & 7.184        & 7.493 &  7.46  & 0.00195\\
    $n_y\rightarrow 5d_{xz}$ & 7.207       & 7.516  &    & 0.00586\\
    \bottomrule
    \end{tabular}
\end{table*}
In the following, we refer to the lines in Figure~\ref{psScan_FEN} and~\ref{Leg2_fen} by their photoelectron energy for the lowest photon energy at a wavelength of \SI{411}{\nano\metre} ($h\nu\approx\SI{3.01}{\electronvolt}$). The line starting at \SI{0.5}{\electronvolt} corresponds to the $3s$ state of fenchone. Fitting by ~\cref{eq:Kastner}, with a $3s$ Rydberg state energy of $E^{3s}_\text{Ryd}= \SI{5.954}{\electronvolt}$ extracted from the \unit{\nano\second} $2+1$ REMPI spectrum, the adiabatic ionization energy was determined to be $I^\text{fen}_{P} = \SI{8.48(3)}{\electronvolt}$ which is consistent with previous work.\cite{Kastner.2017, Singh.2020} Knowing the $I_{P}$, we can fit the other lines with slope one in Figure~\ref{psScan_FEN} by ~\cref{eq:Kastner} to determine the energies $E_\text{Ryd}$ of further Rydberg states, which we do in the following.

We tentatively assign the weak line below \SI{0.5}{\electronvolt} to the $3s$ Rydberg state accompanied by a vibrational excitation in the C--H stretching band \cite{Lee.2022} because the calculations do not hint at any Rydberg states below the $3s$ state. However, the energy difference between this line and the $3s$ line is around \SI{0.36}{\electronvolt}, which matches the energy of the C--H stretch vibration.\cite{Larkin.2017} This notable exception to the $\Delta v =0$ propensity rule is much weaker than the $3s$ line resulting from $\Delta v =0$ and therefore indicates the strength of the propensity.

The photoelectron contributions around \SI{1}{\electronvolt} belong to the $3p$ Rydberg states. The lowest $3p$ state ($3p_{z}$), whose energy from the \unit{\nano\second} spectroscopy is \SI{6.306(7)}{\electronvolt}, is not visible in the photoelectron spectra due to its low two-photon cross-section discussed earlier. The other two $p$ states ($3p_{y}$ and $3p_{x}$) are distinctly visible in the second Legendre coefficient (see  Figure~\ref{Leg2_fen} in the Appendix). For the second $3p$ state ($3p_{y}$), we obtain \SI{6.41(4)}{\electronvolt}, which is in good agreement with the \unit{\nano\second} $2+1$ REMPI assignment and also in accordance with earlier observations.\cite{Kastner.2017,  Kastner.2020, Singh.2020, Powis.2023} For the third $3p$ state ($3p_{x}$), we determine \SI{6.50(5)}{\electronvolt}, which is also in agreement with the \unit{\nano\second} $2+1$ REMPI result of \SI{6.464(7)}{\electronvolt}. This is another confirmation of the $3p$ state assignment performed on the \unit{\nano\second} REMPI spectrum (Figure~\ref{Ns_REMPI_Fen}~(B)).

Note that even at wavelengths longer than the two-photon threshold, photoelectrons whose energies are consistent with $\Delta v=0$ ionization from the $3p$ states are detected, as seen in Figure~\ref{psScan_FEN} and in particular in Figure~\ref{Leg2_fen}. We suspect that two mechanisms can explain these electrons: First, three photons promote an electron from the HOMO--1 orbital to one of the $3p$ orbitals. The HOMO--1 orbital is \SI{1.8}{\eV} more bound than the HOMO orbital.\cite{Powis.2008, Beaulieu.2016b} Then, the molecule is in a doubly-excited state of Rydberg character that has a potential energy surface similar to that of the excited molecular ion with a hole in the HOMO--1 orbital. Another photon can ionize the molecule to this ionic state without changing its vibrational quantum numbers ($\Delta v=0$ propensity). The other possible mechanism exploits that off-resonant states are populated transiently during multi-photon transitions. This transient population vanishes at the end of the laser pulse, but the molecule can be ionized with a low probability during the laser pulse from the transiently populated $3p$ Rydberg states. The energy of the photoelectron will then carry the energy associated with these states.\cite{Krug.2009, Jimenez-Galan.2016} Therefore, in both mechanisms, the ionization step takes place via a Rydberg state, and \cref{eq:Kastner} can be used to determine the state's energy.

Above the photoelectron energy of \SI{1.2}{\electronvolt}, we can see three further lines of photoelectron signals with a slope of one. These feature low count rates because the corresponding higher-lying Rydberg states cannot be reached by two photons at all photon energies. Therefore, these photoelectrons can be attributed to either $3+1$ REMPI, including a hole in the HOMO--1 orbital, or to the ionization of transiently populated states. We find two photoelectron contributions around $\SI{1.5}{\electronvolt}$ which correspond to Rydberg energies of $\SI{6.85(6)}{\electronvolt}$ and $\SI{7.04(6)}{\electronvolt}$. According to the energy difference to the $3s$ state, the TDDFT assigns the lower state as $3d_{xx}$ and the higher state as a mixture of $3d_{yy-zz}$ and $3d_{xz}$ (See Table \ref{fen_TDDFT}). In the CC2 model, the relative energies of these states are about $\SI{.15}{\electronvolt}$ smaller, but we still see this as a confirmation of the TDDFT assignment, considering the general tendency of CC2 to underestimate the energy differences between states. We observe another weak contribution around \SI{2}{\electronvolt}, which corresponds to a Rydberg energy of \SI{7.46(7)}{\electronvolt}. According to the TDDFT model, this is assignable to a $5p$ state. The CC2 calculates close-lying relative energy for the $5p_x$ state, but this does not follow the trend of the CC2 energies always being too low.

Additionally, Figure~\ref{psScan_FEN} contains two lines of photoelectron energies that do not have a slope of one as a function of the photon energy. First, there is a line with a slope of three that starts slightly above the photoelectron energy of \SI{.5}{\electronvolt}. This corresponds to the non-resonant three-photon ionization, where the molecule is directly ionized without passing through an intermediate state. Second, we observe a line around \SI{3.5}{\electronvolt} photoelectron energy, particularly pronounced in Figure~\ref{Leg2_fen}. It has a slope of two, and its photoelectron energies are always one photon energy higher than those corresponding to the $2+1$ REMPI process via the $3s$ Rydberg states. This implies that this high-energy contribution belongs to a $2+2$ resonance-enhanced above-threshold ionization through the same state.\cite{lux.2016,Assion.1997}

\begin{figure}[htbp]
	\begin{center}
		\includegraphics[width=8.4cm]{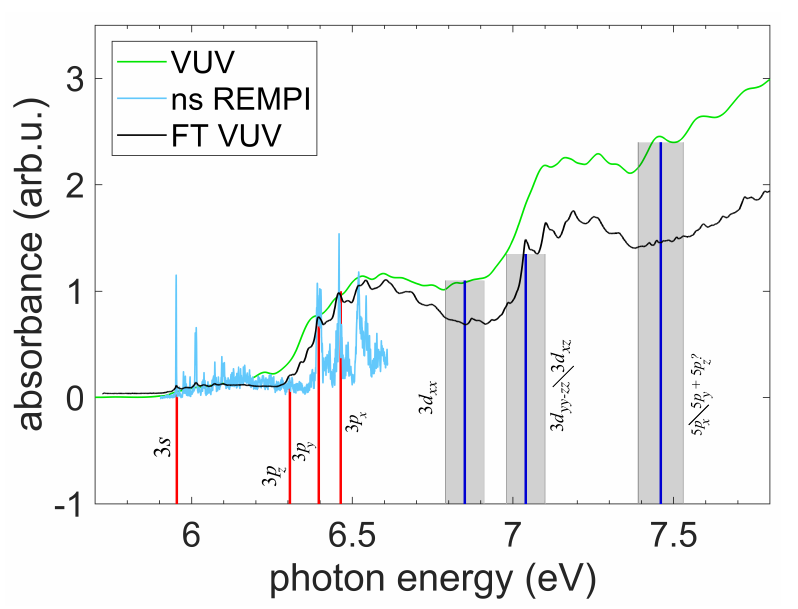}
        	\caption{Single-photon VUV absorption spectrum of gas-phase fenchone measured with a deuterium lamp (green) in comparison with \unit{\nano\second} $2+1$ REMPI (cyan)\cite{Kastner.2020} and FT VUV (black).\cite{Singh.2020} Note that the absorbance scales are individually arbitrary for all spectra. Vertical red lines show the state energies from the \unit{\nano\second} $2+1$ REMPI spectra, where the lines are thicker than the measured uncertainties. Blue vertical lines indicate the state energies from the multi-photon photoelectron spectra with the uncertainties shaded in gray.}
		\label{UV_FEN}
	\end{center}
\end{figure}

In Figure~\ref{UV_FEN}, we present our single-photon VUV absorption spectrum of fenchone (green line). The overall shape is in reasonable agreement with the high-resolution FT-VUV spectrum of Singh \etal (black line).\cite{Singh.2020} We mark the energy values of the $3s$ and $3p$ Rydberg states obtained from the current ns $2+1$ REMPI spectra with red vertical lines. Our energy values agree with both VUV spectra and a high-resolution \unit{\nano\second} REMPI spectrum (cyan line).\cite{Kastner.2020} Furthermore, vertical lines in dark blue mark the energy values of the $3d$ and $5p$ states obtained from the multi-photon photoelectron spectra with the errors shaded in grey. The $3d$ and $5p$ energies are consistent with the band origins in the single-photon absorption spectra within the estimated error (See Figure \ref{UV_FEN}). Note that VUV absorption spectra feature vibrational progressions of the Rydberg states and possibly absorption into non-Rydberg states. The overall shape of the measured single-photon VUV spectrum fits reasonably well with the calculated oscillator strengths $f$ of the TDDFT model (see Table~\ref{fen_TDDFT}). In Figure~\ref{UV_FEN}, the $3s$ Rydberg state has a lower signal than the $3p$ states, the same is true for the calculated $f$ values. However, the calculated oscillator strengths of most of the $3d$ states are weaker than those of the $3p$ states, which only partially agrees with the measured VUV single-photon spectrum.

\subsection{Thiofenchone}
\label{sec:thiofenchone}

\begin{figure}[htbp]
	\begin{center}
		\includegraphics[width=8.2cm]{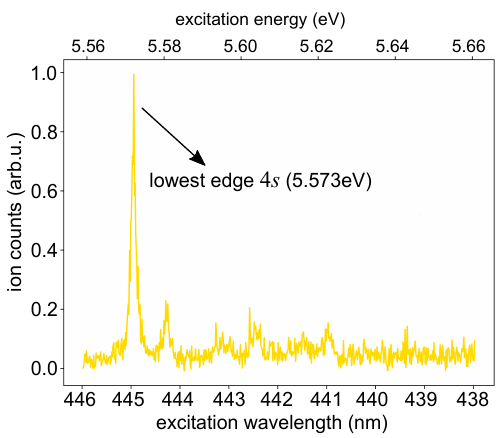}
		\caption{Nanosecond $2+1$ REMPI spectrum of thiofenchone covering the lowest edge of the $4s$ Rydberg state measured with linearly polarized laser pulses.}
		\label{Ns_REMPI_Thiofen}
	\end{center}
\end{figure}

Figure \ref{Ns_REMPI_Thiofen} depicts the \unit{\nano\second} $2+1$ REMPI spectrum measured for thiofenchone. The intense peak at \SI{444.9}{\nano\metre} is the onset of the spectrum and indicates the threshold for reaching the $4s$ Rydberg state. Considering the two-photon character for reaching this resonance, we obtain \SI{5.573(3)}{\electronvolt} as the energy difference between the $4s$ electronic state and the electronic ground state. This compares favorably with the value of \SI{5.55}{\electronvolt}, which was measured at lower resolution in gas phase UV absorption.\cite{Falk.1988} In addition, the measured $4s$ Rydberg state energy is in good agreement with the energy values obtained from the TDDFT ($\SI{5.597}{\electronvolt}$, see Table~\ref{thiofen_TDDFT}) and CC2 ($\SI{5.492}{\electronvolt}$, see Table~\ref{thiofen_CC2}) calculations. The smaller peaks at higher photon energies in the REMPI spectrum of Figure~\ref{Ns_REMPI_Thiofen} are attributed to excited vibrational levels of the $4s$ electronic state.

\begin{figure}[htbp]
	\begin{center}
		\includegraphics[width=8.4cm]{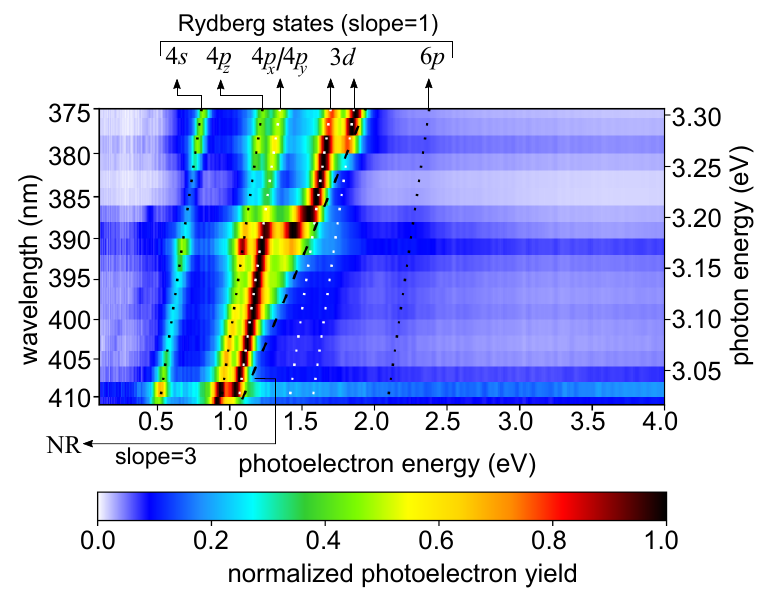}
		\caption{Multi-photon photoelectron spectra of thiofenchone. Each row in the figure represents a measurement at a different central wavelength. All measurements used a pulse duration of \SI{0.4}{\pico\second} and linear polarization. In each row, the signal is normalized to its maximum. Lines indicate linear scalings that can be attributed to different processes: $2+1$ resonance-enhanced multi-photon ionization through the labeled Rydberg states yields a slope of one, while non-resonant (NR) multi-photon ionization is associated with a slope of three. The corresponding second Legendre coefficients are presented in Figure~\ref{Leg2_thiofen} in the appendix.}
		\label{Ps_Thiofen}
	\end{center}
\end{figure}

To determine the adiabatic $I_P$ and to identify more states, we measured multi-photon photoelectron spectra for 19 different central wavelengths ranging from \SIrange{375}{411}{\nano\metre}, which is shown in Figure\ref{Ps_Thiofen}. In analogy to the case of fenchone (Section~\ref{sec:fenchone}), we identify intense lines with a slope of one as associated with Rydberg states via $2+1$ REMPI. The line originating at $\SI{.5}{\electronvolt}$ corresponds to the $4s$ Rydberg state of thiofenchone. Fit by \cref{eq:Kastner}, using the $4s$ Rydberg energy from ns REMPI (\SI{5.573}{\electronvolt}), results in an adiabatic ionization energy $I^\text{thio}_P = \SI{8.07(3)}{\electronvolt}$ which is in close agreement with the vertical $I_P$ of \SI{8.1}{\electronvolt} measured by Sandström \etal\cite{Guimon.1974} and our TDDFT value of $I_P=\SI{8.15}{\electronvolt}$.

\begin{table*}
    \caption{\label{thiofen_TDDFT} TDDFT results for thiofenchone: Vertical electronic singlet excitation energies $E_\text{calc}$, calculated energies $E^*_\text{calc}$ shifted such that the $4s$ energy matches the measured value, experimental state energies $E_\text{exp}$, two-photon excitation cross sections for linearly (circularly) polarized light $\sigma_\text{lin}$ ($\sigma_\text{circ}$) and oscillator strengths $f$. Parentheses mark a tentative assignment.}
    \begin{tabular}{l S[round-mode=places,round-precision=3,group-digits = none] S[round-mode=places,round-precision=3,group-digits = none] c S[round-mode=places,round-precision=4,group-digits = none] S[round-mode=places,round-precision=4,group-digits = none] S[round-mode=places,round-precision=4,group-digits = none]}
    \toprule
    \multicolumn{7}{c}{TDDFT Thiofenchone}\\
    \midrule
    {Electronic Transitions} &{\centering{$E_\text{calc}$ (\unit{\electronvolt})}}&\centering{{$E^*_\text{calc}$ (\unit{\electronvolt})}}&\centering{{$E_\text{exp}$ (\unit{\electronvolt})}}&\centering{{$\sigma \textsubscript{lin}$ ($10^{-50}\unit{\centi\metre\tothe{4}\second}$)}}&\centering{{$\sigma \textsubscript{circ}$ ($10^{-50}\unit{\centi\metre\tothe{4}\second}$)}}&\centering{$f$} \\
    \midrule
    $n_y\rightarrow \pi^*_x$ & 2.635& 2.611&   & 0.00001594 & 0.00002367 & 0.00001592\\
    $\pi_x\rightarrow \pi^*_x$ & 5.517& 5.493  &    & 0.01019978 & 0.00089595 & 0.22317327\\
    $n_y\rightarrow 4s$ & 5.597&5.573 & 5.573 & 0.01821357 & 0.02729300 & 0.02477630\\
    $\sigma_{-2}\rightarrow \pi^*_x$ & 5.969&5.945&  & 0.00063825 & 0.00089693 & 0.01452772\\
    $n_y\rightarrow 4p_z$ & 6.068&6.044& 5.99 & 0.00372915 & 0.00555975 & 0.01289499\\
    $n_y\rightarrow 4p_x$ & 6.157&6.133 & \multirow{2}{*}{6.11} & 0.00754572 & 0.00790320 & 0.00023423\\
    $n_y\rightarrow 4p_y$ & 6.186&6.162&  & 0.00996530 & 0.00203543 & 0.00382284\\
    {$n_y\rightarrow 3d_{xx-zz}$} & 6.461&6.437& 6.46 & 0.01569741 & 0.02340003 & 0.00121326\\
    $\pi_y\rightarrow \pi^*_x$ & 6.481&6.457&  & 0.00156429 & 0.00224350 & 0.00090276\\
    $n_y\rightarrow 3d_{yy}$ & 6.648&6.624& \multirow{2}{*}{6.62} & 0.00292902 & 0.00338011 & 0.00067260\\
    $n_y\rightarrow 3d_{yz}$ & 6.666&6.642&  & 0.09466615 & 0.01453360 & 0.00200265\\
    $n_y\rightarrow 3d_{(xy)}$ & 6.678&6.654&  & 0.02527249 & 0.01135352 & 0.00194947\\
    $n_y\rightarrow 3d_{xz}$ & 6.725&6.701&   & 0.00709153 & 0.01063616 & 0.00044155\\
    $n_y\rightarrow 5s$ & 6.771&6.747&   & 0.00210976 & 0.00184095 & 0.00411748\\
    $\sigma_{-5}\rightarrow \pi^x$ & 6.803&6.779& & 0.00996627 & 0.01198121 & 0.01387741\\
    $n_y\rightarrow 5p_z$ & 6.893&6.869& & 0.00043521 & 0.00028766 & 0.00027396\\
    $n_y\rightarrow 5p_x$ & 6.909&6.885&   & 0.00069600 & 0.00091960 & 0.00002633\\
    $n_y\rightarrow 5p_y$ & 6.928&6.904&   & 0.00421400 & 0.00130143 & 0.00208830\\
    $\sigma_{-4}\rightarrow \pi^*_x$ & 6.949&6.925& & 0.00626264 & 0.00269727 & 0.00265035\\
    \makecell[l]{$n_y\rightarrow 6p_{(z)}$\;(55.59\%),\\$\sigma_{-3}\rightarrow \pi^*_x$\;(43.9\%)} & 7.014&6.99& & 0.00367637 & 0.00441356 & 0.00121426\\
    \makecell[l]{$\sigma_{-3}\rightarrow \pi^*_x$\;(54.72\%),\\$n_y\rightarrow 6p_z$\;(44.61\%)} & 7.022&6.998&  & 0.00232904 & 0.00261378 & 0.01659791\\
    $n_y\rightarrow 6p_x$ & 7.135&7.111& {\multirow{2}{*}{7.14}} & 0.00111657 & 0.00166615 & 0.00054265\\
    $n_y\rightarrow 6p_y$ & 7.156&7.132&  & 0.00801183 & 0.00180565 & 0.00179185\\
    $n_y\rightarrow 4d_{(yz)}+p_y$ & 7.191&7.167&   & 0.00122243 & 0.00045454 & 0.00556486\\
    $n_y\rightarrow 4d_{yy-zz}$ & 7.206&7.182&  & 0.00192809 & 0.00149692 & 0.00325467\\
    $n_y\rightarrow 4d_{xy}$ & 7.211&7.187&  & 0.00184294 & 0.00076107 & 0.00166613\\
    $\pi_x\rightarrow 6s$ & 7.236&7.212&   & 0.00799637 & 0.01054138 & 0.04944759\\
    \bottomrule
    \end{tabular}
\end{table*}

\begin{table*}
    \caption{\label{thiofen_CC2} CC2 results for thiofenchone: Vertical electronic singlet excitation energies $E_\text{calc}$, calculated energies $E^*_\text{calc}$ shifted such that the $4s$ energy matches the measured value, experimental state energies $E_\text{exp}$, and oscillator strengths $f$. Parentheses mark a tentative assignment.}
    \begin{tabular}{c S[round-mode=places,round-precision=3,group-digits = none] S[round-mode=places,round-precision=3,group-digits = none] c S[round-mode=places,round-precision=4,group-digits = none]}
    \toprule
    \multicolumn{5}{c}{CC2 Thiofenchone}\\
    \midrule
    Electronic Transitions &\centering{{$E_\text{calc}$}(\unit{\electronvolt})}&\centering{{$E^*_\text{calc}$}(\unit{\electronvolt})} &\centering{{$E_\text{exp}$}(\unit{\electronvolt})}& \centering{{$f$}} \\
    \midrule
    $n_y\rightarrow \pi^*_x$ & 2.645&2.726 &   & 0.00002\\
    $n_y\rightarrow 4s$ & 5.492&5.573&5.573&   0.01290\\
    $\pi_x\rightarrow \pi^*_x$ & 5.624&5.705 &   & 0.23421\\
    $n_y\rightarrow 4p_z$ & 5.826&5.907& 5.99  & 0.02263\\
    $n_y\rightarrow 4p_x+4p_y$ & 5.958& 6.039 & \multirow{2}{*}{6.11}  & 0.00139\\
    $n_y\rightarrow 4p_y+4p_x$ & 5.976& 6.057  & & 0.00407\\
    $\sigma_{1}\rightarrow \pi^*_x$ & 6.081&6.162&   & 0.02270\\
    $n_y\rightarrow 3d_{xx-zz}$ & 6.282& 6.363   & 6.46  & 0.00089\\
    $n_y\rightarrow 3d_{yz}$ & 6.415&6.496& & 0.00077\\
    $n_y\rightarrow 3d_{yy}$ & 6.427& 6.508   & \multirow{2}{*}{6.62}  & 0.00273\\
    $n_y\rightarrow 3d_{xz}$ & 6.441& 6.522   &   & 0.00093\\
    $n_y\rightarrow 3d_{xy}$ & 6.480&  6.561   &  & 0.00026\\
    $n_y\rightarrow 5s$ & 6.574& 6.655  &  & 0.00511\\
    $\pi_y\rightarrow \pi^*_x$ & 6.671& 6.752   &   & 0.00148\\
    $n_y\rightarrow 5p_z$ & 6.761& 6.842   &  & 0.00385\\
    $n_y\rightarrow 5p_x$ & 6.786&  6.867   &   & 0.00029\\
    $n_y\rightarrow 5p_y$ & 6.800&  6.881   &   & 0.00423\\
    \makecell[l]{$\pi_3\rightarrow \pi^*_x$\;(68.39\%),\\$n_y\rightarrow 6s$\;(31.14\%)} & 6.882& 6.963  &   & 0.00652\\
    \makecell[l]{$n_y\rightarrow 6s$\;(53.58\%),\\$\pi_3\rightarrow \pi^*_x$\;(46.\%)} & 6.886& 6.967  &   & 0.01780\\
    $n_y\rightarrow 4d_{xx-zz}$ & 6.961&  7.042  &   & 0.00326\\
    \bottomrule
    \end{tabular}\\
\end{table*}

With the knowledge of $I_{P}$, we can use fits by ~\cref{eq:Kastner} to the other lines with a slope of one in Figure~\ref{Ps_Thiofen} and Figure~\ref{Leg2_thiofen} to determine the energies $E_\text{Ryd}$ of further Rydberg states of thiofenchone. We attribute the two lines originating near \SI{1}{\electronvolt} to $4p$ states located at $\SI{5.99(5)}{\electronvolt}$ and $\SI{6.11(5)}{\electronvolt}$, respectively. These Rydberg states have also been observed by Falk and Steer,\cite{Falk.1988}, who identified them as $4p_z$ and $4p_y$, respectively. Further evidence for the assignment of the $4p_z$ state at $\SI{5.99(5)}{\electronvolt}$ is based on our calculations: While the energy difference to the $4s$ state from the TDDFT calculation is near the upper limit of the experimental error, the CC2 model yields a lower energy difference of $\SI{.334}{\electronvolt}$, which is close to the experimentally determined error range. The state at $\SI{6.11(5)}{\electronvolt}$ on the other hand may be attributed to either $4p_y$ or $4p_x$, which in the TDDFT calculation have a separation of just $\SI{.03}{\electronvolt}$ - below our experimental resolution. Note that the CC2 model identifies the states as being of mixed $(4p_x + 4p_y)$ character. The calculated two-photon excitation cross-section from the TDDFT model for linearly polarized light is smaller for the $4p_z$ state than for the $4p_x$ and $4p_y$ states, which is in agreement with our observation of the line strength in Figure~\ref{Ps_Thiofen}.

Figure~\ref{Ps_Thiofen} contains two lines originating around \SI{1.5}{\electronvolt}, which show a threshold behaviour near \SI{388}{\nano\metre} and \SI{380}{\nano\metre}, respectively. The energies of the corresponding states are determined to be \SI{6.46(6)}{\electronvolt} and \SI{6.62(6)}{\electronvolt}. Comparison of the energy separation to the $4s$ state with the TDDFT calculation (Table~\ref{thiofen_TDDFT}) suggests that the \SI{6.46(6)}{\electronvolt} state is the lowest-energy $3d$ state, $3d_{xx-zz}$. Considering the tendency of the CC2 model to underestimate the energy differences of the states, this assignment is supported by the CC2 model, too. Similarly, the TDDFT calculation suggests that the $\SI{6.62(6)}{\electronvolt}$ state is $3d_{yz}$ or/and $3d_{yy}$. Of these, the $3d_{yz}$ is much more likely to be populated in two-photon excitation due to the higher cross-section. The multi-photon photoelectron spectra also show a weak line originating at around \SI{2.1}{\eV}, which can be attributed to either 3+1 REMPI, including a hole in the HOMO--1 orbital, or to the ionization of transiently populated non-resonant states. We obtain \SI{7.14(9)}{\eV} for this state, which, according to the TDDFT calculation, can be attributed to the $6p_x$ state with an admixture of the closely-lying $6p_y$ state and various $4d$ states. In the CC2 model, only the energy of the highest calculated state, the $4d_{xx-zz}$, comes close to the measured value.

\begin{figure}[htbp]
	\begin{center}
		\includegraphics[width=8.4cm]{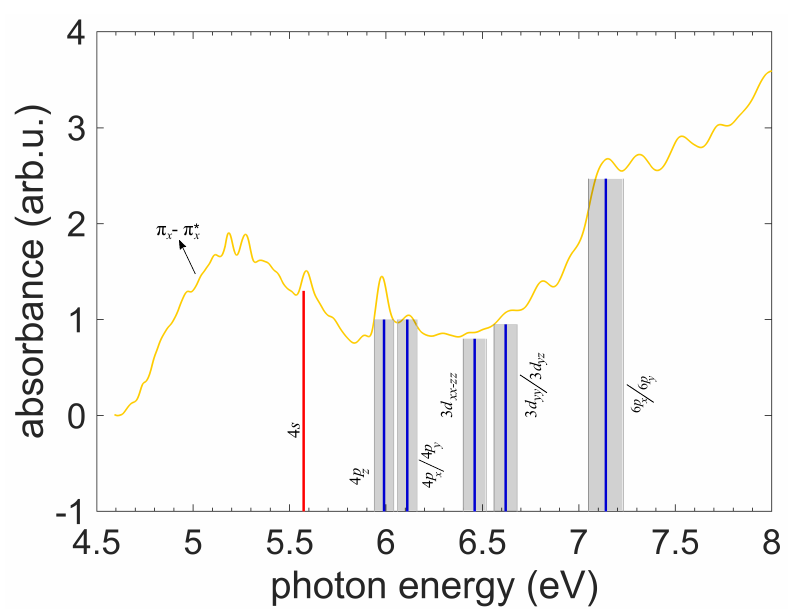}
		\caption{Single-photon VUV absorption spectrum of gas-phase thiofenchone measured with a deuterium lamp. The vertical red line marks the $4s$ state energy from the ns $2+1$ REMPI spectra, where the line is thicker than the measured uncertainty. Blue vertical lines indicate the state energies from the multi-photon photoelectron spectra, with the uncertainties shaded in gray.}
		\label{UV_ThioFEN}
	\end{center}
\end{figure}

In Figure \ref{UV_ThioFEN}, the Rydberg state energies of thiofenchone discussed above are compared with the measured single-photon VUV gas-phase absorption spectrum. Matching absorption peaks for the $4s$ and $4p$ states are found, while the higher-lying Rydberg states are difficult to allocate because possible peaks are indiscernible from the noise (see \nameref{sec:VUV_absorption}). In addition to peaks attributed to Rydberg states, the VUV spectrum exhibits a broad feature starting at \SI{4.6}{eV} and peaking around $\SI{5.2}{\eV}$, which is assigned to the $(\pi_x\rightarrow\pi_x^{*})$ state, which has valence character. For this state, both theories yield vertical transition energies, which are significantly higher than the measured peak value (TDDFT: $\SI{5.517}{\eV}$ and CC2: $\SI{5.624}{\eV}$, i.e. even energetically above the $4s$ Rydberg state). This pronounced deviation may indicate that a detailed vibronic simulation of this broad $(\pi_x\rightarrow\pi_x^{*})$ transition profile would be required to establish the relation between peak maximum and vertical transition energy. Our measured single-photon absorption spectrum reproduces most of the features of the earlier spectrum by Falk and Steer,\cite{Falk.1988} but extends to higher energies. A notable difference is a small modulation on top of the $(\pi_x\rightarrow\pi_x^{*})$ absorption band around $\SI{5.2}{\electronvolt}$, which, however, is attributed to experimental noise. In general, the oscillator strengths $f$ calculated with TDDFT (see Table~\ref{thiofen_TDDFT}) for the $(\pi_x\rightarrow\pi_x^{*})$, $4s$, and $4p$ states are reproduced in the general shape of the UV spectrum, but the model does not find a state with large $f$ that agrees with the strong increase in absorbance above $\SI{6.8}{\eV}$. This increase is likely due to the multitude of states in this energy region. 

\subsection{Selenofenchone}
\label{sec:selenofenchone}

\begin{figure}
	\begin{center}
		\includegraphics[width=8.4cm]{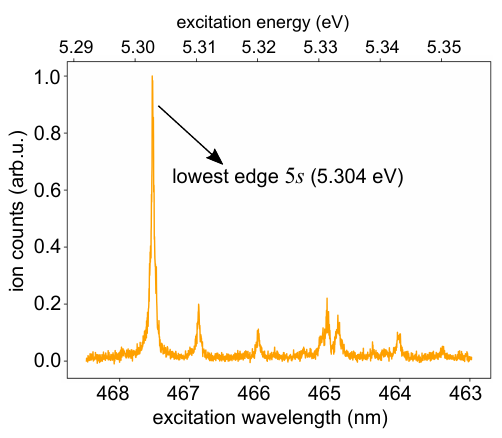}
		\caption{Nanosecond $2+1$ REMPI spectrum of selenofenchone covering the lowest edge of the $5s$ Rydberg state measured with linearly polarized laser pulses.}
		\label{Ns_REMPI_Selenofen}
	\end{center}
\end{figure}

Figure \ref{Ns_REMPI_Selenofen} shows a well-resolved \unit{\nano\second} $2+1$ REMPI spectrum of selenofenchone. The intense peak at \SI{467.5}{\nano\metre} is the onset of the spectrum and indicates the threshold for reaching the $5s$ Rydberg state. Considering the two-photon character for reaching this resonance, we obtain \SI{5.304(3)}{\electronvolt} as the energy difference between the $5s$ Rydberg state and the electronic ground state, which agrees with the calculated TDDFT (Table~\ref{selenofen_TDDFT}) and CC2 energy values (Table~\ref{selenofen_CC2}). The smaller peaks at higher photon energies in the REMPI spectrum of Figure~\ref{Ns_REMPI_Selenofen} are attributed to excited vibrational levels of the $5s$ electronic state. The $5s$ state assigned here could be related to an absorption feature that has previously been observed in different solvents between \num{5.51} and \SI{5.66}{\electronvolt}.\cite{Andersen.1982, Wijekoon.1983} 

\begin{figure}
	\begin{center}
		\includegraphics[width=8.4cm]{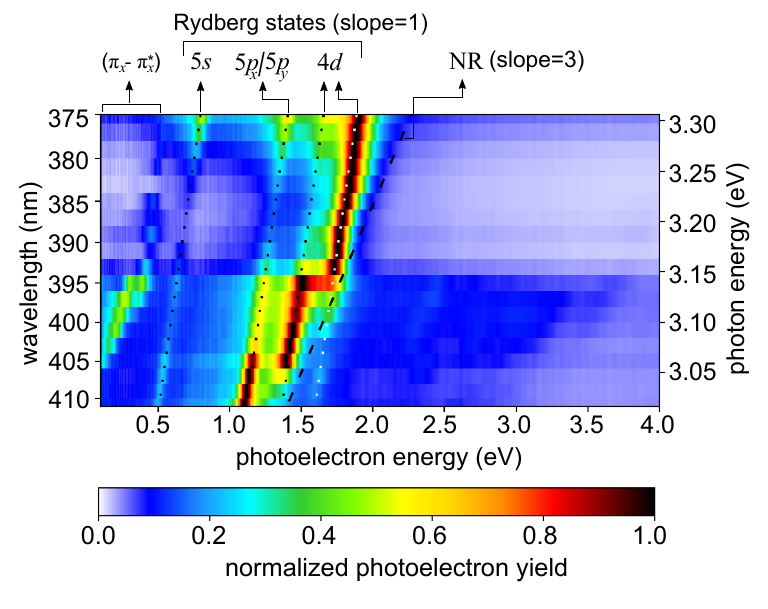}
		\caption{Multi-photon photoelectron spectra of selenofenchone. Each row in the figure represents a measurement at a different central wavelength. All measurements used a pulse duration of \SI{0.4}{\pico\second} and linear polarization. In each row, the signal is normalized to its maximum. Lines indicate linear scalings that can be attributed to different processes: $2+1$ resonance-enhanced multi-photon ionization through the labeled Rydberg states yields a slope of one, while non-resonant (NR) multi-photon ionization is associated with a slope of three. The corresponding second-order Legendre coefficients are presented in Figure~\ref{Leg2_seleno} in the appendix.}
		\label{Ps_selenofen}
	\end{center}
\end{figure}

To determine the adiabatic $I_P$ and to identify more states, we measured multi-photon photoelectron spectra for 19 different central wavelengths ranging from \SIrange{375}{411}{\nano\metre}, which is shown in Figure\ref{Ps_selenofen}. In analogy to the case of fenchone (Section~\ref{sec:fenchone}), we identify lines with a slope of one as associated with Rydberg states via $2+1$ REMPI. The line originating at $\SI{.5}{\electronvolt}$ corresponds to the $5s$ Rydberg state of selenofenchone. Fit by \cref{eq:Kastner}, using the $5s$ Rydberg energy from ns REMPI ($E_\text{Ryd}= \SI{5.304}{\electronvolt}$), results in an adiabatic ionization energy $I^\text{selen}_P = \SI{7.81(3)}{\electronvolt}$ which is in close agreement with the vertical $I_P$ of \SI[round-mode=places,round-precision=2]{7.898}{\electronvolt} determined by the TDDFT calculation.

\begin{table*}
    \caption{\label{selenofen_TDDFT} TDDFT results for selenofenchone: Vertical electronic singlet excitation energies $E_\text{calc}$, calculated energies $E^*_\text{calc}$ shifted such that the $5s$ energy matches the measured value, experimental state energies $E_\text{exp}$, two-photon excitation cross sections for linearly (circularly) polarized light $\sigma_\text{lin}$ ($\sigma_\text{circ}$) and oscillator strengths $f$. Parentheses mark a tentative assignment.}
    \begin{tabular}{l S[round-mode=places,round-precision=3,group-digits = none] S[round-mode=places,round-precision=3,group-digits = none] c S[round-mode=places,round-precision=4,group-digits = none] S[round-mode=places,round-precision=4,group-digits = none] S[round-mode=places,round-precision=4,group-digits = none]}
    \toprule
    \multicolumn{7}{c}{TDDFT Selenofenchone}\\
    \midrule
    {Electronic Transitions} &{\centering{$E_\text{calc}$ (\unit{\electronvolt})}}&\centering{{$E^*_\text{calc}$ (\unit{\electronvolt})}}&\centering{{$E_\text{exp}$ (\unit{\electronvolt})}}&\centering{{$\sigma \textsubscript{lin}$ ($10^{-50}\unit{\centi\metre\tothe{4}\second}$)}}&\centering{{$\sigma \textsubscript{circ}$ ($10^{-50}\unit{\centi\metre\tothe{4}\second}$)}}&\centering{{$f$}} \\
    \midrule
    $n_y\rightarrow \pi^*_x$ & 2.298&2.312&   & 0.00002292 & 0.00003428 & 0.00001788\\
    $\pi_x\rightarrow \pi^*_x$ & 4.866&4.88&   & 0.00817510 & 0.00036781 & 0.23906073\\
    $n_y\rightarrow 5s$ & 5.290&5.304& 5.304 & 0.02353444 & 0.03523823 & 0.03938481\\
    $\sigma_{-2}\rightarrow \pi^*_x$ & 5.445&5.459&  & 0.00083802 & 0.00112587 & 0.00567542\\
    $n_y\rightarrow 5p_z$ & 5.699&5.713&  & 0.00623195 & 0.00930165 & 0.00359416\\
    $n_y\rightarrow 5p_x$ & 5.896&5.91& {\multirow{2}{*}{5.91}} & 0.01716375 & 0.01707154 & 0.00031033\\
    $n_y\rightarrow 5p_y$ & 5.925&5.939&  & 0.04042190 & 0.00967458 & 0.00514730\\
    \makecell[l]{$\pi_y\rightarrow \pi^*_x$\;(84.42\%),\\$n_y\rightarrow 4d_{(xx)}$\;(15.05\%)} & 6.149&6.163&   & 0.00211415 & 0.00292821 & 0.00047078\\
    $n_y\rightarrow 4d_{xx-zz}$ & 6.162&6.176 &6.16 & 0.00772228 & 0.01140159 & 0.00303784\\
    $n_y\rightarrow 4d_{(yy)}$ & 6.348&6.362&  & 0.00546402 & 0.00705801 & 0.00346860\\
    $n_y\rightarrow 4d_{yz+xy}$ & 6.382&6.396& {\multirow{2}{*}{6.39}}& 0.07728493 & 0.01446406 & 0.00522887\\
    $n_y\rightarrow 4d_{xy-yz}$ & 6.391&6.405&   & 0.05114567 & 0.01460259 & 0.00611067\\
    $n_y\rightarrow 4d_{xz}$ & 6.442&6.456&  & 0.00874941 & 0.01302468 & 0.00041955\\
    $n_y\rightarrow 6s$ & 6.487&6.501&   & 0.00109737 & 0.00078363 & 0.00599001\\
    $\sigma_{-3}\rightarrow \pi^*_x$ & 6.536&6.55&  & 0.01593585 & 0.01582766 & 0.01301858\\
    $n_y\rightarrow 6p_z$ & 6.555&6.569&   & 0.00208009 & 0.00303601 & 0.00193013\\
    $n_y\rightarrow 6p_x$ & 6.623&6.637&   & 0.00165320 & 0.00200630 & 0.00004070\\
    $\sigma_{-3}\rightarrow \pi^*_x$ & 6.635&6.649&   & 0.00865732 & 0.00350557 & 0.00009365\\
    $n_y\rightarrow 6p_y$ & 6.642&6.656&   & 0.00623885 & 0.00247428 & 0.00140466\\
    \makecell[l]{$\pi_x\rightarrow 6s$\;(77.7\%),\\$n_y\rightarrow 4f_{zzz}+7p_{(z)}$\;(19.41\%)} & 6.695&6.709&   & 0.00856003 & 0.01223168 & 0.04684669\\
    \makecell[l]{$n_y\rightarrow 4f_{zzz}+7p_z$\;(79.64\%),\\$\pi_x\rightarrow 7s$\;(19.82\%)} & 6.699&6.713&   & 0.00366057 & 0.00546345 & 0.01481280\\
    $\sigma_{-4}\rightarrow \pi^*_x$ & 6.747&6.761&   & 0.01235826 & 0.01315491 & 0.02022651\\
    $n_y\rightarrow 5d_{(xx)}+4f_{xxz-yyz}$ & 6.799&6.813&   & 0.00516162 & 0.00774244 & 0.00108093\\
    $n_y\rightarrow 4f_{(xxx)}+5d_{(xx)}$ & 6.834&6.848&   & 0.00156546 & 0.00233272 & 0.00809901\\
    $n_y\rightarrow 7p_{(x)} + 4f_{(yyy-3xxy)}$ & 6.848&6.862&   & 0.00315748 & 0.00382467 & 0.00186771\\
    $n_y\rightarrow 6d_{yz}$ & 6.862&6.876&   & 0.01217125 & 0.00348785 & 0.00101058\\
    $n_y\rightarrow 6d_{xy}$ & 6.875&6.889&   & 0.00640545 & 0.00607662 & 0.00481633\\
    $n_y\rightarrow 6d_{yy-zz}$ & 6.886&6.900&   & 0.00016549 & 0.00024382 & 0.00245273\\
    $n_y\rightarrow 6d_{(yz)}$ & 6.895&6.909&   & 0.00347462 & 0.00115113 & 0.00024852\\
    $n_y\rightarrow 6d_{xz}$ & 6.926&6.940&   & 0.00374005 & 0.00542251 & 0.00324274\\
    \bottomrule
    \end{tabular}
\end{table*}

With the knowledge of $I_{P}$, we can use fits by \cref{eq:Kastner} to the other lines with a slope of one in Figure~\ref{Ps_selenofen} and Figure~\ref{Leg2_seleno} to determine the energies $E_\text{Ryd}$ of further Rydberg states of selenofenchone. We attribute the line originating around \SI{1}{\electronvolt} to a $5p$ state and obtain its energy \SI{5.91(5)}{\electronvolt}. Around this energy, the TDDFT calculations (See Table \ref{selenofen_TDDFT}), yield $5p_{x}$ and $5p_{y}$ states with an energy separation of just \SI{0.05}{\electronvolt}, below our experimental resolution. We could not observe the lower-lying $5p_z$ state in our multi-photon photoelectron spectra, which is in line with the significantly lower linear two-photon absorption cross-section in the TDDFT calculations. The CC2 model also predicts a small energy separation between the $5p_x$ and the $5p_y$ states, but all $5p$ Rydberg states are $\approx\SI{.1}{\eV}$ lower than the TDDFT values and the measured energy (see Table~\ref{selenofen_CC2}).

\begin{table*}
    \caption{\label{selenofen_CC2} CC2 results for selenofenchone: Vertical electronic singlet excitation energies $E_\text{calc}$, calculated energies $E^*_\text{calc}$ shifted such that the $5s$ energy matches the measured value, experimental state energies $E_\text{exp}$, and oscillator strengths $f$. Parentheses mark a tentative assignment.}
    \begin{tabular}{c S[round-mode=places,round-precision=3,group-digits = none] S[round-mode=places,round-precision=3,group-digits = none] c S[round-mode=places,round-precision=4,group-digits = none]}
    \toprule
    \multicolumn{5}{c}{CC2 Selenofenchone}\\
    \midrule
    Electronic Transitions &\centering{{$E_\text{calc}$}(\unit{\electronvolt})}&\centering{{$E^*_\text{calc}$}(\unit{\electronvolt})} &\centering{{$E_\text{exp}$}(\unit{\electronvolt})}& \centering{{$f$}} \\
    \midrule
    $n_y\rightarrow \pi^*_x$ & 2.228&2.276  & & 0.00002\\
    $\pi_x\rightarrow \pi^*_x$ & 5.042& 5.09  &   & 0.25784\\
    $n_y\rightarrow 5s$ & 5.256&5.304& 5.304 & 0.02675\\
    $n_y\rightarrow 5p_z$ & 5.569&5.617&   & 0.01672\\
    $\sigma_{1}\rightarrow \pi^*_x$ & 5.621& 5.669 &   & 0.01386\\
    $n_y\rightarrow 5p_x$ & 5.756&5.804& {\multirow{2}{*}{5.91}} & 0.00021\\
    $n_y\rightarrow 5p_y$ & 5.777&5.825&   & 0.00287\\
    $n_y\rightarrow 4d_{xx-zz}$ & 6.006&6.054&  6.16 & 0.00056\\
    $n_y\rightarrow 4d_{yz}$ & 6.195&6.243&   & 0.00352\\
    $n_y\rightarrow 4d_{yy}$ & 6.203&6.251&   & 0.00126\\
    $n_y\rightarrow 4d_{yz+xy}$ & 6.231&6.279& {\multirow{2}{*}{6.39}}  & 0.00165\\
    $n_y\rightarrow 4d_{xy-yz}$ & 6.264&6.312&   & 0.00117\\
    $n_y\rightarrow 6s$ & 6.356&6.404&   & 0.00929\\
    $\pi_y\rightarrow \pi^*_x$ & 6.373&6.421&   & 0.00174\\
    $n_y\rightarrow 6p_z$ & 6.497&6.545&   & 0.00055\\
    $n_y\rightarrow 6p_x$ & 6.575&6.623&   & 0.00011\\
    $n_y\rightarrow 6p_y$ & 6.595&6.643&   & 0.00285\\
    $n_y\rightarrow 4f_{zzz}+p_{(z)}$ & 6.616&6.664&   & 0.00162\\
    $\pi_3\rightarrow \pi^*_x$ & 6.633&6.681&   & 0.02339\\
    $\sigma_{2}\rightarrow \pi^*_x$ & 6.708&6.756&   & 0.00038\\
    \bottomrule
    \end{tabular}
\end{table*}

Figure~\ref{Ps_selenofen} contains two lines originating around \SI{1.3}{\electronvolt} and \SI{1.6}{\electronvolt}, which show a threshold behaviour near \SI{388}{\nano\metre} and \SI{379}{\nano\metre}, respectively. Fit of the lower-lying state by~\cref{eq:Kastner} results in an energy of \SI{6.16(6)}{\electronvolt}, which matches the $4d_{xx-zz}$ state in the TDDFT calculation (Table~\ref{selenofen_TDDFT}). The energy determined by the CC2 model is about \SI{.15}{\electronvolt} lower than in the experiment, matching the general trend observed above (Table~\ref{selenofen_CC2}). For the higher-lying state, we obtained the energy of \SI{6.39(6)}{\electronvolt}. At this energy, the TDDFT model calculates the $4d_{yz+xy}$ and $4d_{xy-yz}$ states, which again have a slightly lower energy in the CC2 calculation. Beyond \SI{2}{\electronvolt}, the photoelectron spectra show only weak and featureless contributions, which we are unable to identify.

\begin{figure}[htbp]
	\begin{center}
		\includegraphics[width=8.4cm]{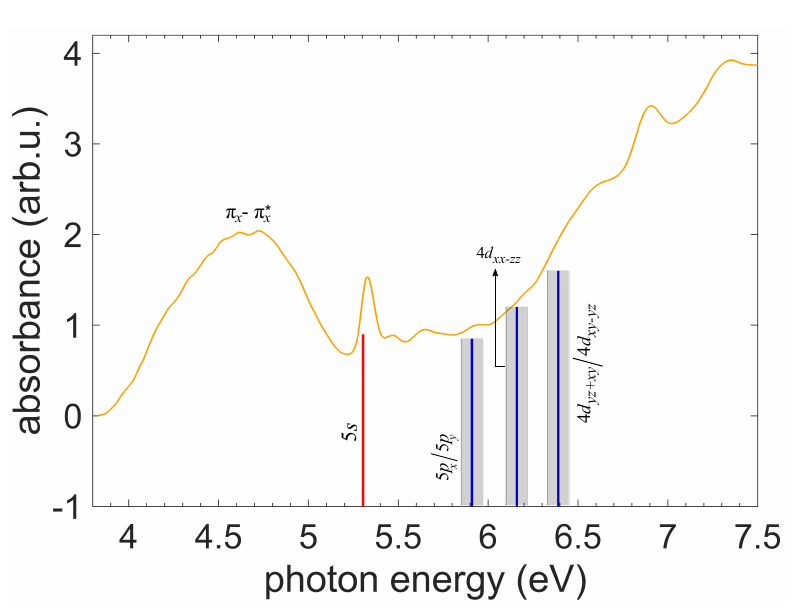}
		\caption{Single-photon VUV absorption spectrum of gas phase selenofenchone measured with a deuterium lamp. The vertical red line shows the $5s$ Rydberg state from ns $2+1$ REMPI spectrum where the energy uncertainty is smaller than the width of the line. Vertical blue lines indicate other Rydberg states from multi-photon photoelectron spectra, with their uncertainties shaded in grey.}
		\label{UV_Selenofen}
	\end{center}
\end{figure}

Aside from the Rydberg states following the $\Delta v=0$ propensity rule, Figure~\ref{Ps_selenofen} features a photoelectron signal below the $5s$ Rydberg state ($< \SI{0.5}{\electronvolt}$). This feature is very prominent in the map of the second-order Legendre coefficient Figure~\ref{Leg2_seleno}. It does not show a linear dependence on photon energy, which implies that its origin is neither non-resonant multi-photon ionization nor REMPI through a Rydberg state. Instead, we suspect a REMPI process via an intermediate ($\pi_x\rightarrow\pi^*_x$) valence-excited state, which is predicted by both our calculations and observed in our single-photon VUV absorption spectrum described below (Figure~\ref{UV_Selenofen}). Interestingly, while the $\pi_x\rightarrow\pi^*_x$ state is seen in the UV absorption spectrum of thiofenchone (see Figure~\ref{UV_ThioFEN}) and should also exist for fenchone, ionization via this state is not observed in the multi-photon photoelectron spectra of these two molecules.

In Figure~\ref{UV_Selenofen}, the determined Rydberg state energies are compared with the measured single-photon VUV gas phase absorption spectrum of selenofenchone. The broad feature starting around $\SI{3.9}{\electronvolt}$ and peaking around $\SI{4.6}{\electronvolt}$ can be assigned to the valence-excited $\pi_x\rightarrow\pi_x^*$ state that previously has been reported to display an absorption maximum at similar photon energies in various solvents.\cite{Wijekoon.1983,Andersen.1982} As in the case of thiofenchone, the vertical transition energies obtained from the calculations (TDDFT: $\SI{4.866}{\electronvolt}$ and CC2: $\SI{5.042}{\electronvolt}$, see Tables~\ref{selenofen_TDDFT} and ~\ref{selenofen_CC2}) overestimate the peak value around $\SI{4.6}{\electronvolt}$. The narrow peak at about \SI{5.3}{\electronvolt} matches the $5s$ Rydberg state. Above this energy there are no clear peaks in the spectrum. However, the absorption gradually increases from about $\SI{5.8}{\electronvolt}$. The previously assigned $5p$ and $4d$ states are in this region of increasing absorption, for which our calculations predict the contributions of multiple states.

\subsection{Scaling of excited state energies in chalcogenofenchones}
\label{sec:overallTrends}

\begin{table*}
\caption{Experimentally determined electronic state energies $E_\text{exp}$ and ionization energies $I_P$ from spectroscopic experiments and the TDDFT assignments of electronic states for fenchone, thiofenchone, and selenofenchone. Parentheses mark a tentative assignment.}
\label{tab:ComparisonExperimental}
\begin{tabular}{>{$}l<{$}S>{$}l<{$}S>{$}l<{$}S}
\toprule
\multicolumn{2}{c}{Fenchone}&\multicolumn{2}{c}{\centering{Thiofenchone}}&\multicolumn{2}{c}{\centering{Selenofenchone}}\\
\cmidrule(r{.5em}){1-2}\cmidrule(lr{.5em}){3-4}\cmidrule(l{.5em}){5-6}
$Electronic States$ & {$E_\text{exp}$ ($\si{\eV}$)} & $Electronic States$ & {$E_\text{exp}$ ($\si{\eV}$)} & $Electronic States$ & {$E_\text{exp}$ ($\si{\eV})$}  \\
\midrule
3s & 5.954 & 4s & 5.573 & 5s& 5.304 \\
3p_{z} & 6.306  & 4p_{z} & 5.99 & & \\
3p_{y} & 6.396 & 4p_{x}/4p_{y} & 6.11 & 5p_{x}/5p_{y} & 5.91\\
3p_{x} & 6.464  & & & & \\
3d_{xx} & 6.85& 3d_{xx-zz} & 6.46 &  4d_{xx-zz} & 6.16 \\
3d_{yy-zz}/3d_{xz} & 7.04 & 3d_{yy}/3d_{yz} & 6.62 & 4d_{yz+xy}/4d_{xy-yz} & 6.39  \\
5p_{x}/5p_{y} + 5p_{(z)}  &7.46 & 6p_{x}/6p_{y} & 7.14 &  & \\
\midrule
I_P & 8.48 & I_P & 8.07 & I_P & 7.81 \\
\bottomrule
\end{tabular}
\end{table*}

\begin{figure}
	\begin{center}
		\includegraphics[width=8.4cm]{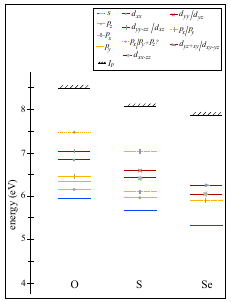}\\
		\caption{Experimentally determined Rydberg state energies and ionization energies $I_P$ of fenchone (O), thiofenchone (S), and selenofenchone (Se). The line colors refer to the angular momentum of the Rydberg state (blue: $l=0$, yellow: $l=1$, red: $l=2$).}
		\label{Exp_Energies}
	\end{center}
\end{figure}
We inspect general trends in the measured and calculated excited state energies of chalcogenofenchones. Table~\ref{tab:ComparisonExperimental} and Figure~\ref{Exp_Energies} summarize the measured Rydberg state energies and ionization energies in fenchone, thiofenchone and selenofenchone. For all three molecules, we could assign at least one $n_\text{min}\,l$ Rydberg state for each angular momentum quantum number $l=\{0,1,2\}$. Here, $n_\text{min}$ is the minimum accessible principal quantum number for each molecule and angular momentum. We see a general trend of bathochromic (red) shifts: The state energies decrease with increasing atomic number of the chalcogen. 

The bathochromic shifts are slightly different depending on the angular momentum of the Rydberg states. For the $n_\text{min}\,s$ states the shift from fenchone to selenofenchone is \SI{.65}{\electronvolt}, while it is only \SI{.49}{\electronvolt} for the $n_\text{min}\,p_y$ states. For the $n_\text{min}\,d$ states, direct comparison is difficult because equivalent orbitals were only found for thiofenchone and selenofenchone. But here for the $n_\text{min}\,d_{xx-zz}$, we get a difference of \SI{.3}{\electronvolt}, which is larger than the shifts between thiofenchone and selenofenchone for the $s$ and $p$ states. In addition, the difference between the ionization energies of fenchone and selenofenchone is \SI{.67}{\electronvolt}, comparable to the shift of the $s$ Rydberg states.

\begin{figure*}
    \includegraphics[width=17.2cm]{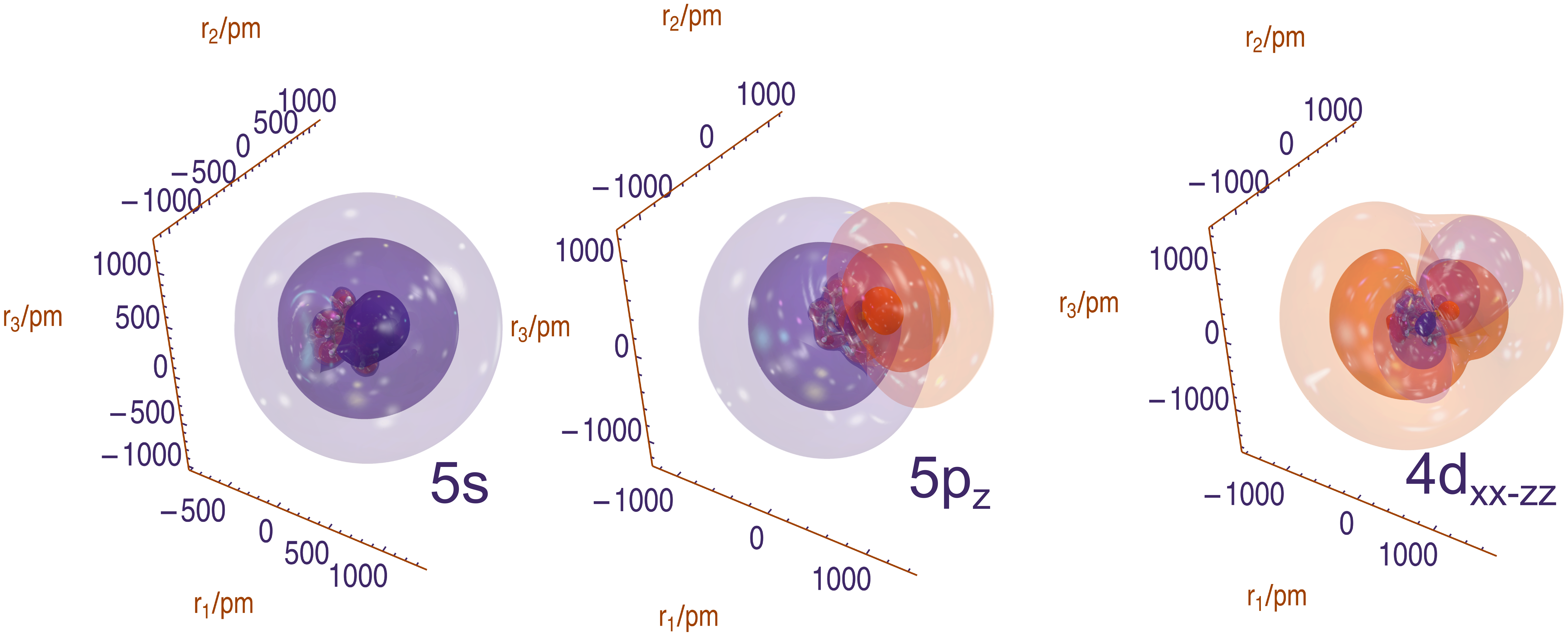}
    \caption{Example electron NTOs for selenofenchone with \(5s\)-, \(5p_z\)- and \(4d_{xx-zz}\)-character,
    i.e. for the lowest energy excitations of each given \(l\),
    showing the influence of the core region on the Rydberg-like NTO. Various intercalated surfaces shown correspond display different isovalues of the NTOs.
    \label{fig:exampleSeVirNTOS}}
\end{figure*}

The trend of $l$-dependent bathochromic shifts is well-reproduced by both calculations when considering the shifted energy values $E_\text{calc}^*$. Angular-momentum-dependent shifts are expected due to the contributions of the molecular core to the Rydberg-like state. The electron NTOs depicted in
Figure~\ref{fig:exampleSeVirNTOS} for the lowest energy \(s,p,d\)
-like states in selenofenchone illustrate this influence, showing regions of different phase of the orbital close to the nuclei and following the shape of the molecule instead of being purely hydrogen-like. Additionally, the figure shows that the  \(4d_{xx-zz}\) orbital of selenofenchone
has a significant admixture of an \(s\)-function, which would further increase the influence of the structure of the molecular system due to the increased probability of the electron being in the core region.

\begin{figure*}
    \includegraphics[width=14cm]{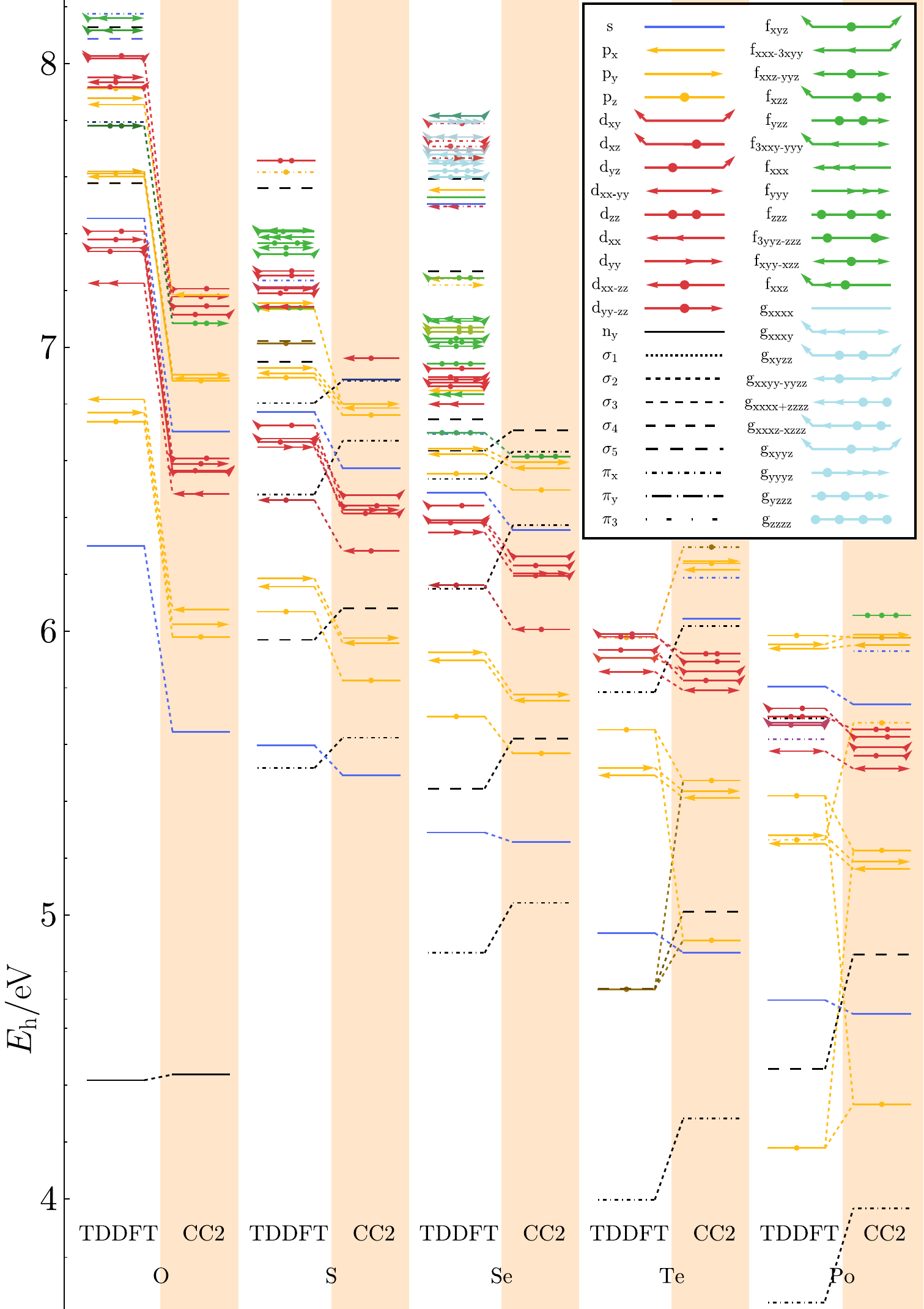}
    \caption{Energetic levels of excitations in the chalcogenofenchones from O to Po with symbolic indication of \(l,m\) quantum-number related characteristics of the Rydberg-like natural transition orbitals. TDDFT-level results are compared to those at CC2 level, with lines connecting states with equivalent characteristics. The colors and symbols are explained in the legend on the top right.}
    \label{fig:ComparisonTheory}
\end{figure*}

For the experimentally identified Rydberg states, the shifting energies as a
function of chemical substitution are reproduced by our theoretical models.
However, they can give a broader overview, as demonstrated in
Figure~\ref{fig:ComparisonTheory}, which includes tellurofenchone and
polonofenchone that were unavailable for the experiment. For these two
molecules, the bathochromic shifts continue. In addition,
Figure~\ref{fig:ComparisonTheory} displays that valence-excited states shift much stronger than Rydberg states. 

Figure~\ref{fig:ComparisonTheory} also helps to compare the two models: For fenchone, the Rydberg-like excitations (non-black in the figure) are strongly reduced in energy in CC2 compared to TDDFT. This is a known failure of the CC2 approximation, which has been shown to give less accurate results for Rydberg-like excitations in small molecules,\cite{Goings.2014}, including
calculations for low-\(l\) excitations in fenchone.\cite{Goetz.2017,Singh.2020}
While the \(n_y\rightarrow \pi^*_x\) valence excitation energies (not shown in the figure) are similar in both models for all chalcogenofenchones, a different trend can be observed for the valence excitations from energetically lower-lying occupied states: While the energy differences between the methods decrease with increasing chalcogen atomic number for the Rydberg-like states,
this difference increases for the valence-excitations not stemming from \(n_y\).

The failure of CC2 for the Rydberg-like states can be attributed to the approximations involved.\cite{Goings.2014} TDDFT, on the other hand, should generally only be used for low-lying valence states,\cite{Laurent.2013} so a larger difference between the methods for higher-energy valence states is not unexpected. However, the fact that the energy differences between the methods for the Rydberg states decrease for the chalcogenofenchone series shows that further benchmarking of the quality of the methods needs to include heavier systems. 

When comparing the performance of the two models to the experiment, we observe that the TDDFT reproduces the energy differences between Rydberg states better than CC2, especially for thiofenchone and selenofenchone.

\section{Femtosecond REMPI-PECD of chalcogenofenchones}
\label{sec:PECD}

\begin{figure*}
	\begin{center}
		\includegraphics[width=17.2cm]{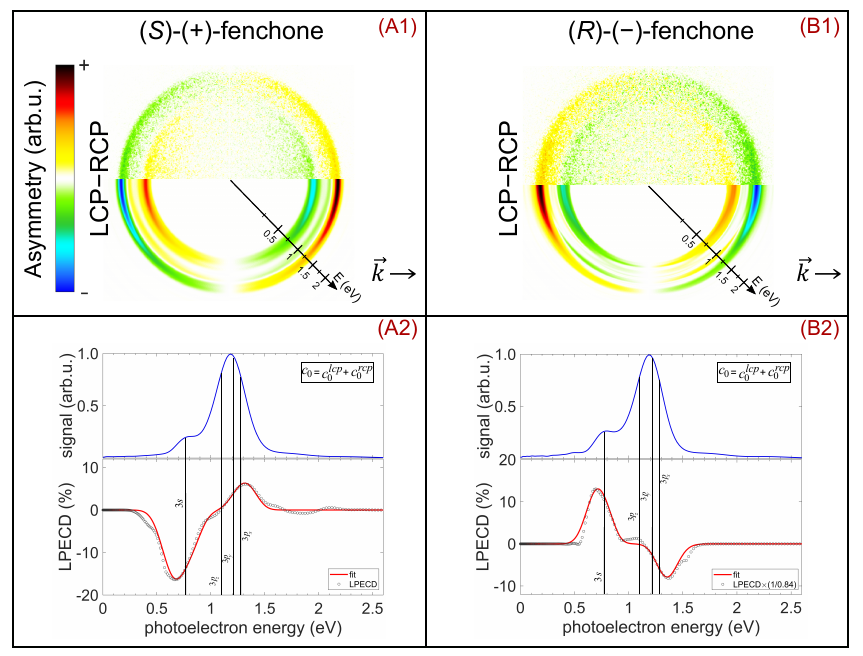}
        
		\caption{Femtosecond $2+1$ REMPI photoelectron circular dichroism of fenchone enantiomers at a central wavelength of \SI{376}{\nano\metre}. In the upper row, (A1) and (B1) show the antisymmetrized PECD images where the laser pulses propagate from left to right. The upper half of each PECD image is raw, and the lower half is Abel inverted. In the lower row, (A2) and (B2) show the total photoelectron signal $c_0$ in blue and the measured and fitted LPECD values as a function of photoelectron energy. The fit is a sum of Gaussian functions. Vertical lines indicate the photoelectron energies corresponding to the Rydberg states that were identified in \nameref{sec:fenchone}.}
		\label{PECD_Fen}
	\end{center}
\end{figure*}

\begin{figure*}
	\begin{center}
		\includegraphics[width=17.2cm]{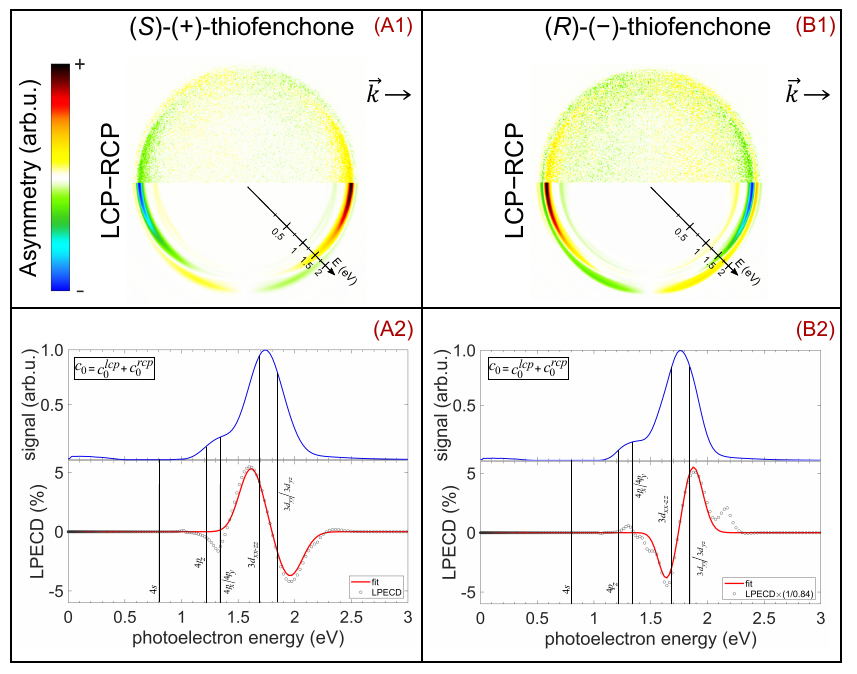}
        
        \caption{Femtosecond $2+1$ REMPI photoelectron circular dichroism of thiofenchone enantiomers at a central wavelength of \SI{376}{\nano\metre}. In the upper row, (A1) and (B1) show the antisymmetrized PECD images where the laser pulses propagate from left to right. The upper half of each PECD image is raw, and the lower half is Abel inverted. In the lower row, (A2) and (B2) show the photoelectron signal $c_0$ in blue and the measured and fitted LPECD values as a function of photoelectron energy. Vertical lines indicate the photoelectron energies corresponding to the Rydberg states that were identified in \nameref{sec:thiofenchone}.}
		\label{PECD_Thiofen}
	\end{center}
\end{figure*}

\begin{figure*}
	\begin{center}
		\includegraphics[width=17.2cm]{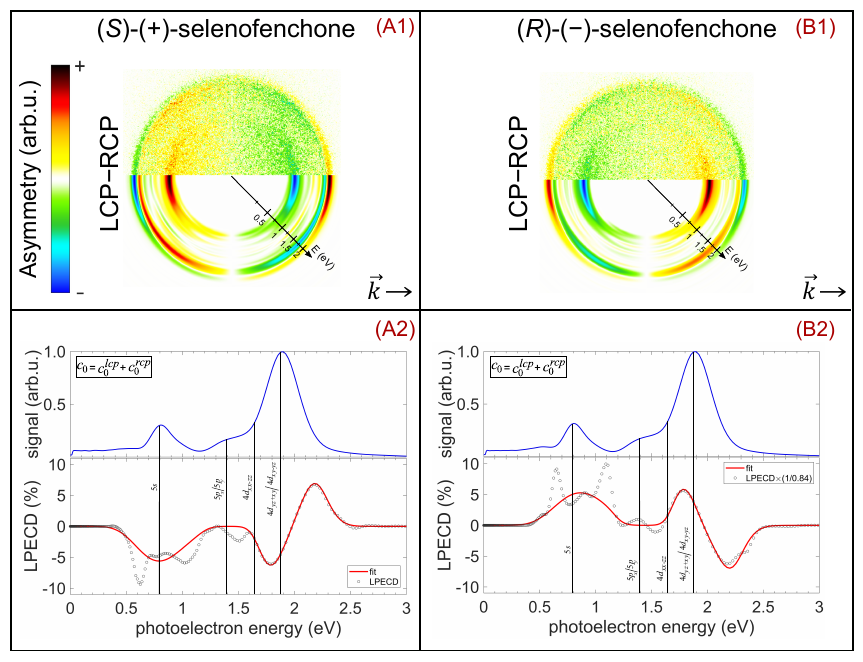}
        
		\caption{Femtosecond $2+1$ REMPI photoelectron circular dichroism results of selenofenchone at a central wavelength of \SI{376}{\nano\metre}. In the upper row, (A1) and (B1) show the antisymmetrized PECD images where the laser pulses propagate from left to right. The upper half of each PECD image is raw, and the lower half is Abel inverted. In the lower row, (A2) and (B2) show the photoelectron signal $c_0$ in blue and the measured and fitted LPECD values as a function of photoelectron energy. Vertical lines indicate the photoelectron energies corresponding to the Rydberg states that were identified in \nameref{sec:selenofenchone}.}
		\label{PECD_selenofen}
	\end{center}
\end{figure*}

In Figures~\ref{PECD_Fen}--\ref{PECD_selenofen} we present experimental femtosecond $2+1$ REMPI-PECD results for both enantiomers of fenchone, thiofenchone, and selenofenchone. All data was obtained at a central wavelength of \SI{376}{\nano\metre}, i.e. very close to the upper end of the multi-photon photoelectron and second-order Legendre coefficient maps presented above. The employed fully compressed, \SI{25}{\femto\second} duration laser pulses minimize the potential influence of molecular dynamics, such as internal conversion and vibration, during the REMPI-PECD process.\cite{Kastner.2017,Lee.2022} The intensity was estimated to \SI{3e12}{\watt\per\centi\metre\squared} for all measurements. In each figure, the results in column (A) were obtained with the (\textit{S})-($+$) enantiomer, while the results in column (B) were obtained with the (\textit{R})-($-$) enantiomer. In the upper row of each figure, panels (A1) and (B1) show the antisymmetrized PECD images with a horizontal laser propagation direction from left to right. The upper half of each PECD image results from the subtraction of raw photoelectron images for left and right circularly polarized light, while the lower half results from the Abel-inverted photoelectron momentum distributions. In the lower row of each figure, panels (A2) and (B2) show the total photoelectron signal $c_0$ as a function of photoelectron energy in blue (upper half). These photoelectron spectra are derived from Abel-inverted photoelectron momentum distributions, which are the sum of the images obtained for the two enantiomers. In the lower half of the lower row, measured LPECD values (open circles) are plotted as a function of photoelectron energy. These graphs are obtained from the odd-order Legendre coefficients of the Abel inversion according to \cref{LPECD}. The blue lines in the photoelectron spectra and the red lines in the LPECD plots represent fits with a sum of Gaussians to the respective data. Vertical lines indicate the photoelectron energies associated with the respective resonant Rydberg states identified in \nameref{sec:spectrocopy}. The direct connection between the PECD images in the upper rows and the LPECD spectra in the lower rows should be noted: Peaks in the latter appear as rings in the former.

First, we compare the photoelectron spectra in panels~(A2) and~(B2) of Figures~\ref{PECD_Fen}--\ref{PECD_selenofen}, measured with \SI{25}{\femto\second} duration laser pulses with circular polarization, with the respective multi-photon photoelectron spectra obtained for linearly polarized pulses with \SI{0.3}{ps} duration, at the same wavelength of \SI{376}{\nano\metre}. For fenchone (Figure~\ref{PECD_Fen}), the signal associated with the $3p_y$ and 3$p_x$ states is about five times more intense than the 3$s$ signal in the femtosecond PES, while the 3$p_y$/3$p_x$ signal is about half of the 3$s$ state signal in the picosecond PES (see Figure~\ref{psScan_FEN}). This behaviour in fenchone has been observed in a recent study of the dependence of PECD on the pulse duration.\cite{Lee.2022} The reason for this observation is a decay of the $3p$ states to the $3s$ state with a lifetime of around \SI{100}{\femto\second}. Hence, the relative yield of resonant ionization via the 3$s$ state is strongly increased for the long pulses compared to the short pulses, which is what we observe comparing Figures~\ref{psScan_FEN} and~\ref{PECD_Fen}. For the same reason, our observation for \SI{25}{fs} pulse duration features a two times stronger relative contribution of 3$p_y$/3$p_x$ as compared to Kastner \etal\cite{Kastner.2017} where the pulse duration was around \SI{55}{fs} for the same wavelength.

For thiofenchone, the ratio of the signal associated with 4$p$ over 3$d$ is about 1:2 in the picosecond photoelectron spectra (see Figure~\ref{Ps_Thiofen}), while it is about 1:5 in the femtosecond photoelectron spectra (see Figure~\ref{PECD_Thiofen}). We conclude that the lifetime of the 3$d$ states is significantly shorter than the lifetime of the 4$p$ states for thiofenchone. The 4$s$ state, due to its presumably long lifetime, is visible in the picosecond photoelectron spectra but not in the femtosecond data of Figure~\ref{PECD_Thiofen}. For selenofenchone, we observe a relative intensity of the $5s$/$5p$/$4d$ signals of 2:3:5 for the picosecond data (Figure~\ref{Ps_selenofen}) and 2:1:6 for the femtosecond data (Figure~\ref{PECD_selenofen}). The constant ratio between $5s$ and $4d$ implies hence that either both of their lifetimes are significantly longer than the \SI{.4}{\pico\second} pulse duration, or that they are both similar. The fact that the relative $5p$ yield is significantly higher for picosecond as compared to femtosecond pulses suggests, however, that $5s$ and $4d$ decay faster than $5p$, on a timescale shorter than the \SI{.4}{\pico\second} pulse duration.

Next, we note that the angle-integrated LPECD values in panels~(A2) and~(B2) of Figures~\ref{PECD_Fen}--\ref{PECD_selenofen} depend on the photoelectron energy and feature both positive and negative signs for all enantiomers and molecules. In some cases, there is a pronounced LPECD effect in the wings of the peaks in the photoelectron spectrum, where the photoelectron yield is already very small, for example, the large LPECD in fenchone associated with the $3s$ state that peaks at an $\approx\SI{.1}{\electronvolt}$ lower energy than the corresponding photoelectron signal (see Figure~\ref{PECD_Fen}~(A2) and~(B2)).

Apart from the different ratio between the $3p$ and $3s$ signals, both our photoelectron and PECD results for fenchone are in good agreement with the earlier work of Kastner \etal\cite{Kastner.2017} The strongest LPECD value at $\approx\pm\SI{15}{\percent}$ is observed for the $3s$ state, while there is a weak LPECD signal of the same sign in the $3p_z$ region. At photoelectron energies closest to the $3p_x$ state, the LPECD peaks at $\approx\mp\SI{7}{\percent}$ with the opposite sign. We cannot exclude that the $3p_y$ state contributes to this LPECD peak as well. We can, however, see in panels~(A1) and~(B1) of Figure~\ref{PECD_Fen} that the two PECD contributions in the $3p$ region feature not only different signs in their LPECD but also different angular distributions: While the PECD for the $3p_x$ and $3p_y$ states has most of its signal at small angles with the light's $k$-vector, the relative contribution at larger angles is higher for the $3p_z$ state. 

For thiofenchone, panels~(A2) and~(B2) of Figure~\ref{PECD_Thiofen} show that the strongest PECD contributions are at photoelectron energies associated with the $3d$ states. For the ($S$)-($+$)-enantiomer, the $3d_{xx-zz}$ exhibits a positive LPECD peaking around \SI{5}{\percent} while the LPECD goes to $\approx\SI{-4}{\percent}$ in the photoelectron energy range associated with the $3d_{yy}$ or $3d_{yz}$ state. In addition, Figure~\ref{PECD_Fen}~(A1) shows remarkably different angular distribution in the $3d$ states: The energetically lower $3d_{xx-zz}$ state mainly exhibits PECD signal close to the propagation direction of the light. In contrast, the $3d_{yy}$ or $3d_{yz}$ state mostly contributes at perpendicular angles. A weak PECD signal with the same sign as the $3d_{yy}$ or $3d_{yz}$ state can be observed in the energy range associated with the $4p$ states. For the ($R$)-($-$)-enantiomer, the signs of all PECD contributions are inverted, as can be seen in panels (B1) and (B2) of Figure~\ref{PECD_Thiofen}.


In selenofenchone, both the $5s$ state and the $4d$ states exhibit significant PECD. For the (\textit{S})-($+$)-enantiomer (Figure~\ref{PECD_selenofen}~(A2)), we observe an LPECD around \SI{-5}{\percent} at the maximum of the $5s$ photoelectron peak and even higher LPECD values at the wings of the peak. Because of the similar redshifts in the lowest $s$ Rydberg state energies and the ionization energies, $3s$ REMPI electrons in fenchone have almost the same energy as $5s$ REMPI electrons in selenofenchone. Nonetheless, the associated $5s$ LPECD in selenofenchone of \SI{5}{\percent} is lower than the \SI{15}{\percent} of the fenchone $3s$ state. 


In the region of the $4d$ Rydberg states, the LPECD of selenofenchone features a sign change. For (\textit{S})-($+$)-se\-le\-no\-fen\-chone, the sign changes from $-$ to $+$ while it changes from $+$ to $-$ in (\textit{S})-($+$)-thiofenchone. A similar difference is found in the angular distribution of the PECD: In (\textit{S})-($+$)-selenofenchone, the lower-energy part has more signal perpendicular to the light's $k$-vector and the higher-energy part is mainly parallel with the light propagation. The reasons for these differences are unknown and call for further investigations. We should also note that only the negative LPECD peak of (\textit{S})-($+$)-selenofenchone coincides with energies where $4d$ states have been identified by our spectroscopic experiments, namely the $4d_{xx-zz}$, $4d_{yz+xy}$ and $4d_{xy-yz}$. For the positive LPECD peak, other Rydberg states, such as $6s$ or $6p_z$, could contribute (cf. Table~\ref{selenofen_TDDFT}). This demonstrates that the differential detection of PECD can be more sensitive to weakly excited states than spectroscopy. For (\textit{R})-($-$)-selenofenchone, the PECD behaves in the same way, but with opposite signs (see panels (A2) and (B2) of Figure~\ref{PECD_selenofen}).

\section{Conclusion and outlook}
\label{Conclusion}
	
With our combination of spectroscopic techniques, we studied the excited states of fenchone, thiofenchone, and selenofenchone. Gas phase spectroscopy was performed for the first time on the latter molecule, identifying one valence-excited and four Rydberg states as well as determining the ionization energy. Other identified states that have not been assigned previously are the $3d$ and $5p$ Rydberg states of thiofenchone and the $5p$ Rydberg state of fenchone. In addition, we improved the resolution in determining the energies of the $4s$ Rydberg state in thiofenchone and the $3p$ Rydberg states in fenchone.

The TDDFT and CC2 calculations supported the state assignment of fenchone, thiofenchone, and selenofenchone. In addition, we have calculated the electronic states of tellurofenchone and polonofenchone. With the increasing ato\-mic number of the chalcogen, we observed different bath\-ochr\-omic shifts in the energies of the electronic states and the ionization energies. The valence-excited (i.e., non\--Ryd\-berg) states showed a stronger bathochromic shift than the Rydberg states, where the amount of shift also depends on the angular momentum quantum number ($l$).

Supplementing the spectroscopic data, we have measured $2+1$ REMPI-PECD using near-ultraviolet \SI{25}{fs} duration laser pulses with a central wavelength of \SI{376}{nm} on fenchone, thiofenchone, and selenofenchone. For all molecules, we could identify the Rydberg states that contribute to the observed PECD signature. The strongest PECD signal in fenchone is attributed to the $3s$ Rydberg state. For thiofenchone, the PECD signal is observed only from $3d$ states, and both $5s$ and $4d$ Rydberg states dominated the REMPI-PECD of selenofenchone. 

We believe this extensive work will pave the way for further gas phase chiral studies and especially coherent control experiments on the series of chalcogenofenchones using a wider choice of light sources extending toward the visible spectral region. 

\clearpage

\section*{Acknowledgements}
We thank Ivan Powis for providing us with the data of hi\-s V\-UV absorption spectrum of fenchone. Most authors acknowledge funding by the Deutsche Fors\-chungs\-gemeinschaft (DFG, German Research Foundation) – Projektnummer 328\-96\-1117 – SFB 1319 ELCH. T. S. acknowledges funding from the Deutsche Forschungsgemeinschaft under grants SCHA 19\-46/5\--1 and INST 186/1302. J.M. acknowledges funding from the German Research Foundation (DFG) within a Heisenberg professorship, reference number 470414645.

\section*{Data availability}
The data that support the findings of this study are available from the corresponding author upon reasonable request.

\section*{Conflict of Interest}

The authors declare no conflict of interest.

\begin{shaded}
\noindent\textsf{\textbf{Keywords:} \keywords} 
\end{shaded}
\newpage

\section{Appendix}    

\subsection{Mass spectra of chalcogenofenchones}

\begin{figure}[tbp]
	\begin{center}
		\includegraphics[width=8.4cm]{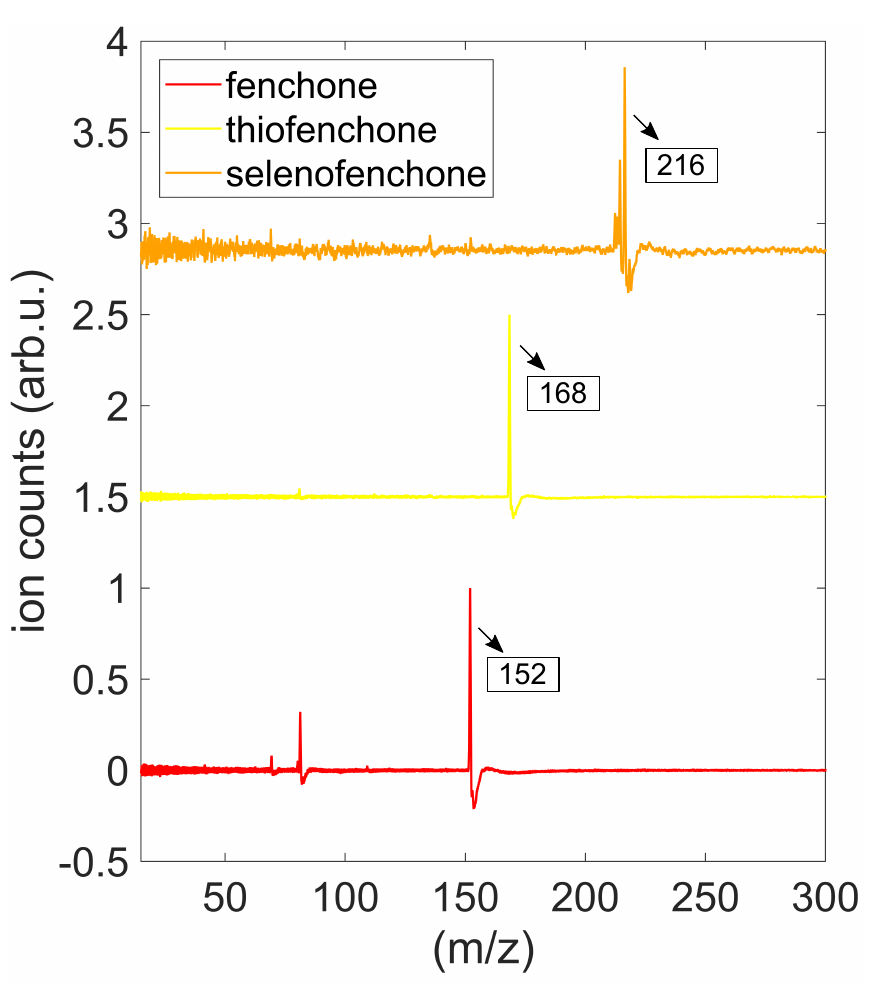}
		\caption{Mass spectra of fenchone, thiofenchone, and selenofenchone measured with linearly polarized \SI{25}{\femto\second} laser pulses centered around \SI{376}{\nano\metre}. The pulse energy was around \SI{3}{\micro\joule} at a repetition rate of \SI{3}{\kilo\hertz}.}
		\label{MassSpec}
	\end{center}
\end{figure}

Figure \ref{MassSpec} displays the mass spectra of fenchone, thiofenchone, and selenofenchone using linearly polarized femtosecond laser pulses with a central wavelength of $\SI{376}{\nano\metre}$. Pulses with the same parameters but circular polarization were used to measure PECD. We observe that the parent ion was dominant for all the molecules. Only for fenchone, a significant amount of mass signals associated with known fragments\cite{Kastner.2019} was found.

\subsection{Second-order Legendre coefficients from multi-photon photoelectron spectroscopy}

\begin{figure}
	\begin{center}
		\includegraphics[width=8.4cm]{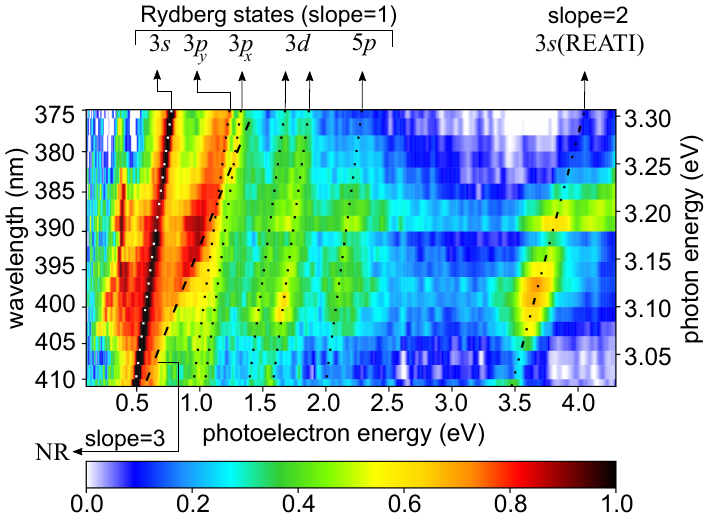}
		\caption{Ratio of second-order Legendre coefficient $c2$ to the total signal $c_0$ for the wavelength-dependent multi-photon photoelectron spectra of fenchone presented in Figure~\ref{psScan_FEN}.}
		\label{Leg2_fen}
	\end{center}
\end{figure}

\begin{figure}
	\begin{center}
		\includegraphics[width=8.4cm]{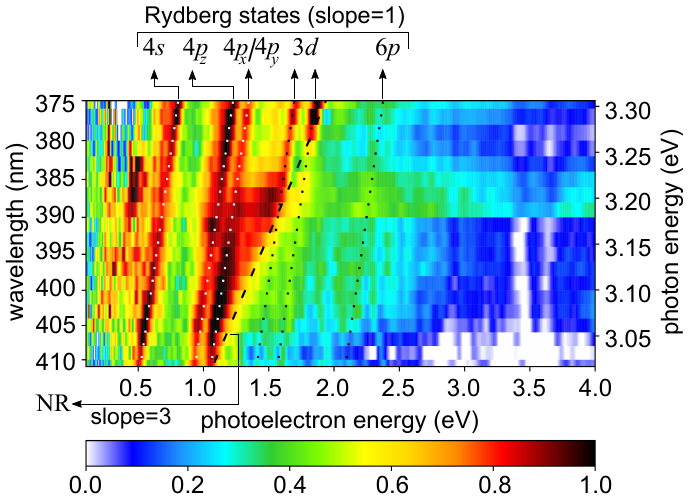}
		\caption{Ratio of second-order Legendre coefficient $c2$ to the total signal $c_0$ for the wavelength-dependent multi-photon photoelectron spectra of thiofenchone presented in Figure~\ref{Ps_Thiofen}.}
		\label{Leg2_thiofen}
	\end{center}
\end{figure}

\begin{figure}
	\begin{center}
		\includegraphics[width=8.4cm]{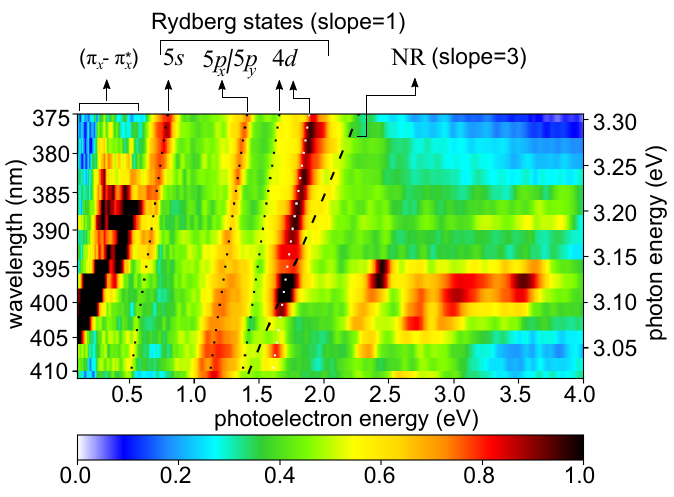}
		\caption{Ratio of second-order Legendre coefficient $c2$ to the total signal $c_0$ for the wavelength-dependent multi-photon photoelectron spectra of selenofenchone presented in Figure~\ref{Ps_selenofen}.}
		\label{Leg2_seleno}
	\end{center}
\end{figure}

We show the second-order Legendre coefficient $c_2$ spectra obtained from multi-photon photoelectron spectroscopy for fenchone (Figure \ref{Leg2_fen}), thiofenchone (Figure \ref{Leg2_thiofen}), and sele\-nofen\-chone (Figure \ref{Leg2_seleno}). In all the second-order Legendre coefficient spectra, each wavelength measurement is normalized by the respective photoelectron spectrum $c_0$.

\newpage

\subsection{Theoretical results for tellurofenchone and polonofenchone}

\begin{table*}
    \caption{\label{TDDFT_tellurofen} TDDFT results for tellurofenchone: Vertical electronic singlet excitation energies $E_\text{calc}$, two-photon excitation cross sections for linearly (circularly) polarized light $\sigma_\text{lin}$ ($\sigma_\text{circ}$) and oscillator strengths $f$. Parentheses mark a tentative assignment.}
    \begin{tabular}{c S[round-mode=places,round-precision=3,group-digits = none]  S[round-mode=places,round-precision=4,group-digits = none] S[round-mode=places,round-precision=4,group-digits = none] S[round-mode=places,round-precision=4,group-digits = none]}
    \toprule
    \multicolumn{5}{c}{TDDFT Tellurofenchone}\\
    \midrule
    {Electronic Transitions} & {$E_\text{calc}$(\unit{\electronvolt})}&{$\sigma \textsubscript{lin}$ ($10^{-50}\unit{\centi\metre\tothe{4}\second}$)}&{$\sigma \textsubscript{circ}$ ($10^{-50}\unit{\centi\metre\tothe{4}\second}$)}& {$f$} \\
    \midrule
    $n_y\rightarrow \pi^*_x$ & 1.823& 0.00003100 & 0.00004650 & 0.00002088\\
    $\pi_x\rightarrow \pi^*_x$ & 3.996& 0.00673516 & 0.00012276 & 0.23275532\\
    \makecell[l]{$n_y\rightarrow 6p_z$\;(59.07\%),\\$\sigma_{-2}\rightarrow \pi^*_x$\;(40.49\%)} & 4.738 & 0.01686633 & 0.02514930 & 0.01474861\\
    $n_y\rightarrow 6s$ & 4.936&0.00637642 & 0.00949494 & 0.03624530\\
    $n_y\rightarrow 6p_x$ & 5.491& 0.03861665 & 0.04019333 & 0.00039009\\
    $n_y\rightarrow 6p_y$ & 5.518 & 0.13290758 & 0.03200337 & 0.00847208\\
    $n_y\rightarrow 6p_z+5d_{zz}$ & 5.652   & 0.00089514 & 0.00132066 & 0.00935570\\
    $\sigma_{-4}\rightarrow \pi^*_x$ & 5.785  & 0.00196076 & 0.00152743 & 0.00045473\\
    $n_y\rightarrow 5d_{xx-yy}$ & 5.857 & 0.02665652 & 0.03970495 & 0.01743095\\
    \makecell[l]{$n_y\rightarrow 5d_{xy}$\;(84.09\%),\\$\pi_x\rightarrow 7p_z$\;(15.51\%)} & 5.905 & 0.02598741 & 0.03573123 & 0.01234274\\
    $n_y\rightarrow 5d_{yz}$ & 5.933& 0.14772478 & 0.02421316 & 0.02021394\\
    \makecell[l]{$\pi_x\rightarrow 7p_z(+s)$\;(81.5\%),\\$n_y\rightarrow 5d_{yy+zz}$\;(18.16\%)} & 5.977& 0.02534200 & 0.03782001 & 0.01215600\\
    $n_y\rightarrow 5d_{zz}$ & 5.980 & 0.00725482 & 0.01087170 & 0.00265463\\
    $n_y\rightarrow 5d_{xz}$ & 5.990 & 0.01081344 & 0.01544870 & 0.00093825\\
    \bottomrule
    \end{tabular}
\end{table*}

\begin{table*}
    \caption{\label{CC2_tellurofen} CC2 results for tellurofenchone: Vertical electronic singlet excitation energies $E_\text{calc}$, and oscillator strengths $f$.}
    \begin{tabular}{c S[round-mode=places,round-precision=3,group-digits = none]  S[round-mode=places,round-precision=4,group-digits = none]}
    \toprule
    \multicolumn{3}{c}{CC2 Tellurofenchone}\\
    \midrule
    {Electronic Transitions} & {$E_\text{calc}$(\unit{\electronvolt})}& {$f$} \\
    \midrule
    $n_y\rightarrow \pi^*_x$ & 1.680  & 0.00002\\
    $\pi_x\rightarrow \pi^*_x$ & 4.282 & 0.24373\\
    $n_y\rightarrow 6s+6p_z$ & 4.867  & 0.04122\\
    $n_y\rightarrow 6p_z+6s$ & 4.911  & 0.00698\\
    $\sigma_{1}\rightarrow \pi^*_x$ & 5.010  & 0.00585\\
    $n_y\rightarrow 6p_x$ & 5.413  & 0.00027\\
    $n_y\rightarrow 6p_y$ & 5.436  & 0.00501\\
    $n_y\rightarrow 6p_z+5d_{zz}$ & 5.473  & 0.01100\\
    $n_y\rightarrow 5d_{xx-yy}$ & 5.791  & 0.01346\\
    $n_y\rightarrow 5d_{yz}$ & 5.826  & 0.01471\\
    $n_y\rightarrow 5d_{xy}$ & 5.859  & 0.00829\\
    $n_y\rightarrow 5d_{xz}$ & 5.893  & 0.00232\\
    $n_y\rightarrow 5d_{zz}$ & 5.920  & 0.00857\\
    $\pi_y\rightarrow \pi^*_x$ & 6.017  & 0.00052\\
    $n_y\rightarrow 7s$ & 6.044  & 0.00626\\
    $\pi_x\rightarrow 6s$ & 6.189  & 0.07910\\
    $n_y\rightarrow 7p_x$ & 6.216  & 0.00113\\
    $n_y\rightarrow 7p_z$ & 6.239  & 0.00618\\
    $n_y\rightarrow 7p_y$ & 6.246  & 0.00333\\
    \bottomrule
    \end{tabular}
\end{table*}

\begin{table*}
    \caption{\label{TDDFT_polonofen} TDDFT results for polonofenchone: Vertical electronic singlet excitation energies $E_\text{calc}$, two-photon excitation cross sections for linearly (circularly) polarized light $\sigma_\text{lin}$ ($\sigma_\text{circ}$) and oscillator strengths $f$.}
    \begin{tabular}{c S[round-mode=places,round-precision=3,group-digits = none]  S[round-mode=places,round-precision=4,group-digits = none] S[round-mode=places,round-precision=4,group-digits = none] S[round-mode=places,round-precision=4,group-digits = none]}
    \toprule
    \multicolumn{5}{c}{TDDFT Polonofenchone}\\
    \midrule
    {Electronic Transitions} & {$E_\text{calc}$(\unit{\electronvolt})}&{$\sigma \textsubscript{lin}$ ($10^{-50}\unit{\centi\metre\tothe{4}\second}$)
}&{$\sigma \textsubscript{circ}$ ($10^{-50}\unit{\centi\metre\tothe{4}\second}$)
}& {$f$} \\
    \midrule
    $n_y\rightarrow \pi^*_x$ & 1.674 & 0.00005555 & 0.00008329 & 0.00001951\\
    $\pi_x\rightarrow \pi^*_x$ & 3.64& 0.00658182 & 0.00003311 & 0.20970841\\
    $n_y\rightarrow 7p_z$ & 4.179& 0.02604951 & 0.03907127 & 0.02361157\\
    $\sigma_{-2}\rightarrow \pi^*_x$ & 4.457& 0.00160851 & 0.00197983 & 0.00088704\\
    $n_y\rightarrow 7s$ & 4.700 & 0.00763529 & 0.01139838 & 0.04066455\\
    $n_y\rightarrow 7p_x$ & 5.251& 0.05789511 & 0.06868893 & 0.00025883\\
    $\pi_x\rightarrow 7p_z$ & 5.264& 0.04055890 & 0.05783772 & 0.01472076\\
    $n_y\rightarrow 7p_y$ & 5.280& 0.22287750 & 0.05796395 & 0.00894687\\
    $n_y\rightarrow 7p_z+6d_{zz}$ & 5.420& 0.00378066 & 0.00565800 & 0.00902478\\
    $n_y\rightarrow 6d_{xx-yy}$ & 5.577& 0.05101135 & 0.07651589 & 0.01634819\\
    \makecell[l]{$\pi_x\rightarrow 8s$\;(55.41\%),\\$n_y\rightarrow 6d_{xy}$\;(44.28\%)} & 5.619& 0.00500642 & 0.00666633 & 0.01549885\\
    \makecell[l]{$n_y\rightarrow 6d_{yz+xy}$\;(83.06\%),\\$\pi_x\rightarrow 8s$\;(16.46\%)} & 5.669& 0.14449660 & 0.03562533 & 0.04386865\\
    \makecell[l]{$n_y\rightarrow 6d_{xy-yz}$\;(74.43\%),\\$\pi_x\rightarrow 8s$\;(24.66\%)} & 5.677& 0.06725406 & 0.03839741 & 0.05978012\\
    $\sigma_{-4}\rightarrow \pi^*_x$ & 5.692& 0.00511403 & 0.00496470 & 0.00150751\\
    $n_y\rightarrow 6d_{zz}$ & 5.698& 0.00481688 & 0.00709584 & 0.00623484\\
    $n_y\rightarrow 6d_{xz}$ & 5.728& 0.01447718 & 0.02045232 & 0.00064059\\
    $n_y\rightarrow 8s$ & 5.805& 0.00519711 & 0.00737244 & 0.01450817\\
    $n_y\rightarrow 8p_x$ & 5.938& 0.00547226 & 0.00796589 & 0.00004516\\
    $n_y\rightarrow 8p_y$ & 5.953& 0.01795995 & 0.01100791 & 0.00165809\\
    $n_y\rightarrow 8p_z$ & 5.984& 0.00561763 & 0.00842569 & 0.00423833\\
    \bottomrule
    \end{tabular}
\end{table*}

\begin{table*}
    \caption{\label{CC2_polonofen} CC2 results for polonofenchone: Vertical electronic singlet excitation energies $E_\text{calc}$, and oscillator strengths $f$.}
    \begin{tabular}{c S[round-mode=places,round-precision=3,group-digits = none]  S[round-mode=places,round-precision=4,group-digits = none]}
    \toprule
    \multicolumn{3}{c}{CC2 Polonofenchone}\\
    \midrule
    {Electronic Transitions} & {$E_\text{calc}$(\unit{\electronvolt})}& {$f$} \\
    \midrule
    $n_y\rightarrow \pi^*_x$ & 1.454  & 0.00002\\
    $\pi_x\rightarrow \pi^*_x$ & 3.966  & 0.21998\\
    $n_y\rightarrow 7p_z+7s$ & 4.332  & 0.03979\\
    $n_y\rightarrow 7s+7p_z$ & 4.651  & 0.02596\\
    $\sigma_{1}\rightarrow \pi^*_x$ & 4.860  & 0.00253\\
    $n_y\rightarrow 7p_x$ & 5.161  & 0.00015\\
    $n_y\rightarrow 7p_y$ & 5.187  & 0.00471\\
    $n_y\rightarrow 7p_z+6d_{zz}$ & 5.227  & 0.01146\\
    $n_y\rightarrow 6d_{xx-yy}$ & 5.515  & 0.01847\\
    $n_y\rightarrow 6d_{yz}$ & 5.561  & 0.01903\\
    $n_y\rightarrow 6d_{xy}$ & 5.591  & 0.01585\\
    $n_y\rightarrow 6d_{xz}$ & 5.627  & 0.00222\\
    $n_y\rightarrow 6d_{zz}$ & 5.654  & 0.00458\\
    $\pi_x\rightarrow 7p_z+7s$ & 5.676  & 0.04492\\
    $n_y\rightarrow 8s+6d_{zz}$ & 5.741  & 0.01463\\
    $\pi_x\rightarrow 7s+7p_z$ & 5.929  & 0.03889\\
    $n_y\rightarrow 8p_x$ & 5.950  & 0.00222\\
    $n_y\rightarrow 8p_z$ & 5.977  & 0.00711\\
    $n_y\rightarrow 8p_y$ & 5.987  & 0.00268\\
    $n_y\rightarrow 5f_{zzz}$ & 6.055  & 0.01664\\
    \bottomrule
    \end{tabular}
\end{table*}

\newpage


\setlength{\bibsep}{0.0cm}
\bibliographystyle{Wiley-chemistry}
\bibliography{Reference}

\clearpage








\noindent\rule{11cm}{2pt}
\begin{minipage}{11cm}
\includegraphics[width=11cm]{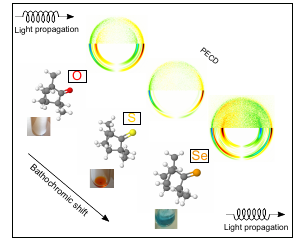}
\end{minipage}
\begin{minipage}{11cm}
\large\textsf{Fenchone, thiofenchone, and selenofenchone (chalcogenofenchones) are characterized using gas phase spectroscopic experiments. The experimentally determined energies of excited states are supported by quantum chemical calculations. Furthermore, near-ultraviolet femtosecond laser pulses were used to measure the photoelectron circular dichroism of both enantiomers.}
\end{minipage}
\noindent\rule{11cm}{2pt}


\end{document}